\documentclass[ALICE,manyauthors]{cernphprep}
\usepackage[comma,square,numbers,sort&compress]{natbib}
\usepackage{hyperref}
\usepackage{lineno}
\usepackage{xspace}
\usepackage{color}
\usepackage{float}
\usepackage[T1]{fontenc}
\begin{document}
%

\newcommand{\pp}           {pp\xspace}
\newcommand{\ppbar}        {\mbox{$\mathrm {p\overline{p}}$}\xspace}
\newcommand{\XeXe}         {\mbox{Xe--Xe}\xspace}
\newcommand{\PbPb}         {\mbox{Pb--Pb}\xspace}
\newcommand{\pA}           {\mbox{pA}\xspace}
\newcommand{\pPb}          {\mbox{p--Pb}\xspace}
\newcommand{\AuAu}         {\mbox{Au--Au}\xspace}
\newcommand{\dAu}          {\mbox{d--Au}\xspace}
\newcommand{\CuCu}         {\mbox{Cu--Cu}\xspace}

\newcommand{\s}            {\ensuremath{\sqrt{s}}\xspace}
\newcommand{\snn}          {\ensuremath{\sqrt{s_{\mathrm{NN}}}}\xspace}
\newcommand{\pt}           {\ensuremath{p_{\rm T}}\xspace}
\newcommand{\meanpt}       {$\langle p_{\mathrm{T}}\rangle$\xspace}
\newcommand{\ycms}         {\ensuremath{y_{\rm CMS}}\xspace}
\newcommand{\ylab}         {\ensuremath{y_{\rm lab}}\xspace}
\newcommand{\etarange}[1]  {\mbox{$\left | \eta \right |~<~#1$}}
\newcommand{\yrange}[1]    {\mbox{$\left | y \right |~<$~0.5}}
\newcommand{\dndy}         {\ensuremath{\mathrm{d}N_\mathrm{ch}/\mathrm{d}y}\xspace}
\newcommand{\dndeta}       {\ensuremath{\mathrm{d}N_\mathrm{ch}/\mathrm{d}\eta}\xspace}
\newcommand{\avdndeta}     {\ensuremath{\langle\dndeta\rangle}\xspace}
\newcommand{\dNdy}           {\ensuremath{\mathrm{d}N_\mathrm{ch}/\mathrm{d}y}\xspace}
\newcommand{\dNdyy}         {\ensuremath{\mathrm{d}N/\mathrm{d}y}\xspace}
\newcommand{\Npart}        {\ensuremath{N_\mathrm{part}}\xspace}
\newcommand{\Ncoll}        {\ensuremath{N_\mathrm{coll}}\xspace}
\newcommand{\dEdx}         {\ensuremath{\textrm{d}E/\textrm{d}x}\xspace}
\newcommand{\RpPb}         {\ensuremath{R_{\rm pPb}}\xspace}
\newcommand{\RAA}         {\ensuremath{R_{\rm AA}}\xspace}

\newcommand{\nineH}        {$\sqrt{s}~=~0.9$~Te\kern-.1emV\xspace}
\newcommand{\seven}        {$\sqrt{s}~=~7$~Te\kern-.1emV\xspace}
\newcommand{\twoH}         {$\sqrt{s}~=~0.2$~Te\kern-.1emV\xspace}
\newcommand{\twosevensix}  {$\sqrt{s}~=~2.76$~Te\kern-.1emV\xspace}
\newcommand{\five}         {$\sqrt{s}~=~5.02$~Te\kern-.1emV\xspace}
\newcommand{\twosevensixnn}{$\sqrt{s_{\mathrm{NN}}}~=~2.76$~Te\kern-.1emV\xspace}
\newcommand{\fivenn}       {$\sqrt{s_{\mathrm{NN}}}~=~5.02$~Te\kern-.1emV\xspace}
\newcommand{\LT}           {L{\'e}vy-Tsallis\xspace}
\newcommand{\GeVc}         {Ge\kern-.1emV$/c$\xspace}
\newcommand{\MeVc}         {Me\kern-.1emV$/c$\xspace}
\newcommand{\TeV}          {Te\kern-.1emV\xspace}
\newcommand{\GeV}          {Ge\kern-.1emV\xspace}
\newcommand{\MeV}          {Me\kern-.1emV\xspace}
\newcommand{\GeVmass}      {Ge\kern-.2emV$/c^2$\xspace}
\newcommand{\MeVmass}      {Me\kern-.2emV$/c^2$\xspace}
\newcommand{\lumi}         {\ensuremath{\mathcal{L}}\xspace}

\newcommand{\ITS}          {\rm{ITS}\xspace}
\newcommand{\TOF}          {\rm{TOF}\xspace}
\newcommand{\ZDC}          {\rm{ZDC}\xspace}
\newcommand{\ZDCs}         {\rm{ZDCs}\xspace}
\newcommand{\ZNA}          {\rm{ZNA}\xspace}
\newcommand{\ZNC}          {\rm{ZNC}\xspace}
\newcommand{\SPD}          {\rm{SPD}\xspace}
\newcommand{\SDD}          {\rm{SDD}\xspace}
\newcommand{\SSD}          {\rm{SSD}\xspace}
\newcommand{\TPC}          {\rm{TPC}\xspace}
\newcommand{\TRD}          {\rm{TRD}\xspace}
\newcommand{\VZERO}        {\rm{V0}\xspace}
\newcommand{\VZEROA}       {\rm{V0A}\xspace}
\newcommand{\VZEROC}       {\rm{V0C}\xspace}
\newcommand{\Vdecay} 	   {\ensuremath{V^{0}}\xspace}

\newcommand{\ee}           {\ensuremath{e^{+}e^{-}}} 
\newcommand{\pip}          {\ensuremath{\pi^{+}}\xspace}
\newcommand{\pim}          {\ensuremath{\pi^{-}}\xspace}
\newcommand{\kap}          {\ensuremath{\rm{K}^{+}}\xspace}
\newcommand{\kam}          {\ensuremath{\rm{K}^{-}}\xspace}
\newcommand{\pbar}         {\ensuremath{\rm\overline{p}}\xspace}
\newcommand{\kzero}        {\ensuremath{{\rm K}^{0}_{\rm{S}}}\xspace}
\newcommand{\lmb}          {\ensuremath{\Lambda}\xspace}
\newcommand{\almb}         {\ensuremath{\overline{\Lambda}}\xspace}
\newcommand{\Om}           {\ensuremath{\Omega^-}\xspace}
\newcommand{\Mo}           {\ensuremath{\overline{\Omega}^+}\xspace}
\newcommand{\X}            {\ensuremath{\Xi^-}\xspace}
\newcommand{\Ix}           {\ensuremath{\overline{\Xi}^+}\xspace}
\newcommand{\Xis}          {\ensuremath{\Xi^{\pm}}\xspace}
\newcommand{\Oms}          {\ensuremath{\Omega^{\pm}}\xspace}
\newcommand{\degree}       {\ensuremath{^{\rm o}}\xspace}
\newcommand{\kstar}        {\ensuremath{\rm {K}^{\rm{* 0}}}\xspace}
\newcommand{\phim}        {\ensuremath{\phi}\xspace}
\newcommand{\pik}          {\ensuremath{\pi\rm{K}}\xspace}
\newcommand{\kk}          {\ensuremath{\rm{K}\rm{K}}\xspace}
\newcommand{\kskm}{$\mathrm{K^{*0}/K^{-}}$}
\newcommand{\phikm}{$\mathrm{\phi/K^{-}}$}
\newcommand{\phixi}{$\mathrm{\phi/\Xi}$}
\newcommand{\phiom}{$\mathrm{\phi/\Omega}$}
\newcommand{\xiphi}{$\mathrm{\Xi/\phi}$}
\newcommand{\omphi}{$\mathrm{\Omega/\phi}$}
\newcommand{\kstf} {K$^{*}(892)^{0}~$}
\newcommand{\phf} {$\mathrm{\phi(1020)}~$}
\newcommand{\dd}{\ensuremath{\mathrm{d}}}
\newcommand{\mT}{\ensuremath{m_{\mathrm{T}}}\xspace}
\newcommand{\krr}{\ensuremath{\kern-0.09em}}

\begin{titlepage}
\PHyear{2021}       
\PHnumber{200}      
\PHdate{09 September}  

\title{$\mathrm{K}^{*}(\mathrm{892})^{0}$ and $\mathrm{\phi(1020)}$  production in p--Pb collisions at \snn  = 8.16 TeV }
\ShortTitle{$\mathrm{K}^{*}(\mathrm{892})^{0}$ and $\mathrm{\phi(1020)}$  in p--Pb at \snn  = 8.16 TeV }   

\Collaboration{ALICE Collaboration\thanks{See Appendix~\ref{app:collab} for the list of collaboration members}}
\ShortAuthor{ALICE Collaboration} 
\begin{abstract}
  The production of \kstf and \phf resonances has been measured in ~p--Pb collisions at  \snn  = 8.16 TeV  using the   ALICE detector.
    Resonances are reconstructed via their hadronic decay channels in the rapidity interval 
    $-$0.5 $<$ $y$ $<$ 0 and the transverse momentum spectra are measured for various multiplicity classes 
    up to \pt  = 20 \GeVc  for \kstf and \pt  = 16 \GeVc for $\mathrm{\phi(1020)}$. 
    The \pt-integrated yields and mean transverse momenta are reported and compared with  previous results in pp, p--Pb and Pb--Pb collisions.
    The $x_{\mathrm{T}}$ scaling for \kstf and \phf resonance production is newly 
    tested in \pPb collisions and found to hold in the  high--\pt region at Large Hadron Collider energies. 
      The nuclear modification factors (\RpPb) as a function of \pt for \kstar and \phim 
     at \mbox{\snn  = 8.16 TeV} are presented along with the new \RpPb measurements of \kstar, \phim,  $\Xi$, and $\Omega$      
     at \mbox{\snn  = 5.02 TeV}. At  intermediate \pt \mbox{(2--8 \GeVc)}, \RpPb of  $\Xi$, $\Omega$  show  a   
     Cronin-like enhancement, while \kstar and \phim show no or little nuclear modification. At high \pt ($>$ 8 \GeVc), the  \RpPb values of all hadrons  are consistent with unity within uncertainties. The  \RpPb of \kstf and \phf at \snn  = 8.16 and 5.02 TeV show  no significant energy dependence.  
     
\end{abstract}
\end{titlepage}

\setcounter{page}{2} 


\section{Introduction}
 High-energy heavy-ion (A--A) collisions provide a unique opportunity to study the deconfined  quark--gluon plasma (QGP) created in such collisions~\cite{Braun-Munzinger:2015hba, Adams:2005dq,Schukraft:2011na}. The hot and dense medium created in heavy-ion collisions evolves with time and cools down to form a phase where hadron resonance gas is studied. Evidence at Relativistic Heavy Ion Collider (RHIC) and the Large Hadron Collider (LHC) suggest that the QGP phase is followed by a hadronic phase where the hadrons interact via rescattering and regeneration processes,  before the final freeze-out. 
Resonances are short-lived hadrons that decay via the strong interactions. They play an important role to understand the particle production mechanisms and for the characterization of the dynamic evolution of the system formed in heavy-ion collisions. They are used as  a sensitive probe of the hadronic phase, where their mass, width and yield could be modified due to interaction of their decay products through re-scattering and regeneration processes ~\cite{Aggarwal:2010mt, Adams:2004ep, Adler:2002sw, Anticic:2011zr, Abelev:2008zk, Abelev:2008aa, Adamczyk:2015lvo, Alt:2008iv, Adare:2010pt,Abelev:2014uua, Adare:2014eyu, Adam:2017zbf}. \mbox{ALICE} has previously measured \kstf and \phf production in pp collisions at $\sqrt{s}$  = 5.02, 7, 8 and 13 TeV~\cite{Abelev:2012hy,ALICE:2017jyt,Acharya:2019qge,ALICE:2021ptz,Acharya:2018orn,Acharya:2019wyb,Acharya:2019bli}, in \pPb collisions at  \snn  = 5.02 TeV~\cite{Adam:2016bpr} and Pb--Pb collisions at  \snn  = 2.76  and 5.02 TeV~\cite{Abelev:2014uua,Adam:2017zbf,Acharya:2019qge,ALICE:2021ptz}.   
  
Proton--lead collisions are intermediate between pp and Pb--Pb collisions 
in terms of the size of the colliding system and the produced particle multiplicities. Recent measurements in high--multiplicity pp, \pPb and d--Au collisions at different energies have uncovered strong flow-like effects even in these small collision systems~\cite{Abelev:2012sk,Acharya:2019bli, Adare:2014eyu,Abelev:2008yz,Adam:2016bpr}, whose origin is not fully understood. 
To investigate the mechanism of  particle production and the origin of these effects, the ALICE Collaboration has studied the multiplicity dependence of light--flavor particle production for many species like $\pi^{\pm}$, K$^{\pm}$, K$_{\mathrm{S}}^{0}$, K$^{*}(892)^{0}$, $\mathrm{\phi(1020)}$, $\Lambda$, $\Lambda(1520)$, $\Sigma^{*\pm}$, $\Xi^{\pm}$, $\Xi^{*0}$, $\Omega^{\pm}$ in \pPb collisions at  \snn  = 5.02 TeV ~\cite{Adam:2016dau,Adam:2015vsf,Adam:2016bpr,ALICE:2019smg,ALICE:2017pgw} and in pp collisions at $\sqrt{s}$  = 7 and 13 TeV~\cite{Acharya:2018orn,Acharya:2019bli,ALICE:2017jyt,ALICE:2014zxz}. This paper reports on the multiplicity dependence  of \kstf and \phf meson production at the highest center-of-mass energy,  \snn  = 8.16 TeV, reached at the LHC in \pPb collisions. This provides an opportunity to extend the previous studies of production of these particles in \pPb collisions at  \snn  = 5.02 TeV~\cite{Adam:2016bpr} to a higher multiplicity reach and a larger \pt coverage.
Hadron production is governed by the soft and hard scattering processes at LHC energies. 
The bulk of particles produced in high energy collisions is dominated by low transverse momentum particles from soft interactions, which are nonperturbative in nature. The yield of particles at low \pt is not well understood from the first principles of QCD and their description relies on phenomenological QCD-based models such as EPOS-LHC, DPMJET and HIJING. The measurements in the low-momentum region of the spectra presented in this article provide input for the tuning of these event generators.
In this paper, measurements of \kstf and $\mathrm{\phi(1020)}$ are compared with predictions from EPOS-LHC~\cite{Pierog:2013ria}, DPMJET~\cite{Roesler:2000he} and HIJING~\cite{Gyulassy:1994ew}.

The transverse momentum spectra of light--flavor hadrons have shown a clear evolution with multiplicity in high-energy pp and \pPb collisions~\cite{ALICE:2017jyt,Acharya:2019bli,Adam:2016bpr,Acharya:2019kyh,Adam:2016dau}, similar to that observed in \PbPb collisions~\cite{Abelev:2013vea,Acharya:2019yoi, Acharya:2019qge,ALICE:2021ptz,Abelev:2014uua}, where in the latter case the effect is usually attributed to a collective expansion of the system. 
The increase in slope of the \pt  spectra  as a function of multiplicity attributed to the radial flow is related to the  low--\pt region of the spectrum, where flow is relevant. This feature is also reflected in an increase of the average transverse momentum \meanpt with multiplicity. In contrast to the yields  \dNdyy, which evolve smoothly as a function
of multiplicity for different collision systems, the  \meanpt  of light--flavour hadrons as well as \kstf and $\mathrm{\phi(1020)}$, rises faster as a function of multiplicity in pp and p--Pb collisions than in Pb--Pb collisions, as discussed in Refs.~\cite{Acharya:2018orn,Adam:2016bpr,Acharya:2019bli}. The new measurements, with the highest multiplicity reach in \pPb collisions, and comparison with  the different model predictions  can be used to further extend these studies.

The high--\pt particle production is analyzed within the framework of  perturbative Quantum Chromodynamics (pQCD) which features a nearly  scale-invariant  behavior of elementary parton--parton hard-scattering processes~\cite{Collins:1989gx,deFlorian:2014xna}. 
The convolution of hard scattering cross sections with the parton
distribution functions (PDFs) of incident hadrons and fragmentation functions (FFs) leads to the observed scaling of the inclusive invariant cross section $E\mathrm{d}^{3}\sigma /\mathrm{d}p^{3}$ 
as $p_{\rm{T}}^{-n}$ at fixed transverse $x$, \mbox{$x_{\rm T}$ $=$ 2\pt$/\sqrt{s}$}~\cite{Brodsky:2005fza,Arleo:2009ch}.
The exponent  $n$ can be related to the scattering processes in which high--\pt hadrons are produced. If hadrons are produced by leading twist (LT) $2 \rightarrow 2$ hard subprocesses, then $\rm{n}$ $\approx$ 4, and for higher twist (HT) processes, $n$ $\approx$ 8. 
 It has been observed that the exponent value decreases with increasing collision energy, which suggests that the contribution of higher twist processes on high--\pt hadron production is reduced as a function of energy.  
 The transverse momentum distributions of different particle species at high \pt are observed to satisfy a universal $x_{\rm{T}}$ scaling over a wide energy range up to
  $\sqrt{s}$  = 13 TeV. This scaling behavior was observed by the CDF~\cite{Aaltonen:2009ne,CDF:2001hmt,Abe:1988yu} and  UA1~\cite{Albajar:1989an} Collaborations in p($\overline{\rm p}$) collisions, and by the STAR~\cite{Adams:2006nd},  ALICE~\cite{ALICE:2020jsh} and CMS~\cite{Chatrchyan:2011av} Collaborations in pp collisions. In this paper, the $x_{\rm{T}}$ scaling of \kstf and \phf mesons are tested in \pPb collisions at  LHC energies. 
The transverse momentum distributions of the particles in \pPb collisions are compared to those in pp collisions  using the nuclear modification factor (\RpPb). The  measurement of  \RpPb acts as a control experiment observable in \pPb collisions~\cite{Acharya:2018qsh} in the context of the observed  high--\pt hadron suppression in \PbPb collisions~\cite{Adam:2017zbf}.   
In this paper,  \RpPb measurements of \kstf and \phf in p--Pb collisions at \snn  = 5.02 and 8.16 TeV, and that of $\Xi$ and $\Omega$ in \pPb collisions at 5.02 TeV are reported.
  Similar measurements are also reported for strange and multi-strange hadrons by CMS~\cite{Sirunyan:2019rfz}, and for $\pi^{\pm}$, K$^{\pm}$ and p($\overline{\rm p}$) by ALICE~\cite{Adam:2016dau} in \pPb collisions at \snn  = 5.02 TeV.
  At  high \pt ($>$ 8 \GeVc), the values of  \RpPb  for all light hadrons are similar and found to be consistent with unity within the uncertainties. At intermediate \pt (2 $<$ \pt  $<$ 8 \GeVc), the  values of \RpPb  for strange baryons ($\Xi$, $\Omega$) show an enhancement with a clear mass dependence~\cite{Sirunyan:2019rfz}. In this \pt region, the hard scattering processes start to dominate over soft processes and the momentum range where this transition may happen depend on the mass and quark composition of the particle species. The measurements of strange particles produced in high multiplicity p--Pb collisions~\cite{Adam:2015vsf,CMS:2016zzh} suggested the presence of radial flow~\cite{ALICE:2013wgn}. Due to the radial flow effect, hadrons of greater mass are pushed towards the higher transverse momentum and the effect increases with hadron mass as well as multiplicity~\cite{ALICE:2013wgn,Adam:2016bpr}. However, it should be noted that some final state effects such as color reconnection in PYTHIA~\cite{Sjostrand:2014zea} which can mimic the radial flow-like effect and EPOS-LHC~\cite{Pierog:2013ria}which uses parameterized flow could describe the modification of transverse momentum spectra. The difference in the production mechanism of baryon and meson has been observed in particle ratios~\cite{ALICE:2013wgn,Adam:2016bpr} and the nuclear modification factors~\cite{Abelev:2008yz,Adams:2006nd,Adam:2016dau,Sirunyan:2019rfz,ALICE:2021est}. The enhanced production of baryon (\RpPb $>$ 1) may happen as a result of hadronization by parton recombination~\cite{Fries:2003vb}.
In addition, there are several initial-state effects such as isospin effect,
Cronin effect, cold--nuclear matter energy loss and nuclear shadowing
that can result in \RpPb $\neq$ 1~\cite{Kang:2012kc}. The Cronin enhancement~\cite{Kopeliovich:2002yh} in the intermediate \pt are reported in the low energy experiments~\cite{Cronin:1973fd,Cronin:1974zm}. Similar enhancement is observed for (anti-)proton compared to pion and kaon in p--Pb collisions at \snn $=$ 5.02 TeV~\cite{Adam:2016dau}. In this paper, the particle species and collision energy dependence of  \RpPb is studied for p--Pb collisions at  LHC energies. \\

  Throughout this paper, the results for K$^{*}(892)^{0}$ and $\mathrm{\overline{K}}^{*}(892)^{0}$ are averaged 
   and denoted by the symbol \kstar, while \phf is denoted by  \phim. The paper is organized as follows.  In Sec. 2, the dataset, event and track selection criteria, the analysis techniques, the procedure for extraction of the yields, and  the study of the systematic uncertainties are briefly discussed. In Sec. 3, the results on the transverse momentum spectra,  the \dNdyy, \meanpt, $x_{\rm{T}}$ scaling, and  \RpPb  in \pPb collisions at  \snn  = 8.16 TeV are presented. Finally, the results are summarized in Sec. 4.

\section{Data analysis}
The measurements of  \kstar and \phim meson production in \pPb  collisions at  \snn  = 8.16 TeV have been performed on data collected with the ALICE detector in the year 2016. The resonances are reconstructed via their hadronic decay channels with branching ratios (BR) of 66.6$\%$ for $\kstar \rightarrow\pi^{\pm}K^{\mp}$ and 49.2$\%$ for \mbox{\phim
$\rightarrow$ K$^{+}$K$^{-}$} in the rapidity interval $-$0.5 $<$ $y$ $<$ 0, where $y$ stands for the rapidity in the nucleon--nucleon center-of-mass. For both \kstar and \phim, the analysis is performed in various multiplicity classes and also using a multiplicity-integrated sample.

\subsection{Event selection} 
The detailed description of the ALICE detector setup and its performance can be found in Refs.~\cite{ Aamodt:2008zz,Abelev:2014ffa}.
In \pPb configurations, the ${}^{208}$Pb beam circulates towards the  positive $z$ direction in the ALICE laboratory frame, while the proton beam circulates in the opposite direction. Due to the asymmetric system, the center-of-mass frame is shifted in the rapidity by $\Delta y$ = --0.465 in the direction of the proton beam with respect to the laboratory frame. 
The  minimum bias trigger was configured to select events by requiring at least a coincidence signal in both the V0A and V0C detectors~\cite{Abbas:2013taa,Acharya:2018egz}. The V0 detector system consists of two arrays of 32 scintillator detectors, one on each side of the interaction point covering the full azimuthal angle in the pseudorapidity regions 
2.8 $<$ $\eta$ $<$ 5.1  (V0A) and \mbox{--3.7 $<$ $\eta$ $<$ --1.7 (V0C)}. 
The background events due to beam--gas interaction and other machine-induced background collisions are rejected using the timing information from the V0 and the Zero Degree Calorimeter (ZDC)~\cite{Abelev:2014ffa}.
The primary vertex of a collision is determined using charged tracks reconstructed in the Inner Tracking System (ITS)~\cite{Aamodt:2010aa} and the Time Projection Chamber (TPC)~\cite{Alme:2010ke}.
The events are selected whose primary vertex position along the beam axis (v$_{z}$, $z$ is the longitudinal direction) is  within $\pm$ 10 cm from the nominal interaction point. 
Pile-up events from the triggered bunch crossing are rejected if multiple collision vertices are identified in the Silicon Pixel Detector (SPD), which is the  innermost detector of the ITS~\cite{Abelev:2014ffa,Alme:2010ke}. The total number of events analyzed after applying the event selection criteria is about 30 million. The minimum bias events are further divided into seven multiplicity classes, according to the total charge deposited in the forward V0A detector~\cite{Abbas:2013taa}. The yield of \kstar and \phim are measured in the rapidity interval$-$0.5 $< y <  $0 for the following event multiplicity classes, 0--5$\%$, 5--10$\%$, 10--20$\%$, 20--40$\%$, 40--60$\%$, 60--80$\%$ and 80--100$\%$. 
The \pt spectra  normalized to  the fraction of non-single-diffractive (NSD) events are also obtained for both \kstar and \phim. The mean charged-particle multiplicity 
$(\langle \mathrm{d}N_{\mathrm{ch}}/\mathrm{d}\eta\rangle)$ corresponding to each multiplicity class, and  measured in the pseudorapidity interval $|\eta_{\mathrm{lab}}|$~$<$~0.5, is given in Table~\ref{tab:chMult} taken from~\cite{Acharya:2018egz}.
\begin{table}
\begin{center}
	\caption{Mean charged particle multiplicity densities $(\langle \mathrm{d}N_{\mathrm{ch}}/\mathrm{d}\eta \rangle)$ measured in pseudorapidity range $|\eta_{\mathrm{lab}}|<$ 0.5, corresponding to the various multiplicity classes defined using the V0A detector in p--Pb collisions at \snn  = 8.16 TeV ~\cite{Acharya:2018egz}. }
    \label{tab:chMult}	
\begin{tabular}{l l}
\hline
\hline
 V0A percentile $(\%)$  &   $\langle \mathrm{d}N_{\mathrm{ch}}/\mathrm{d}\eta\rangle_{|\eta_{\mathrm{lab}}|< 0.5}$  \\
\hline 
 0--5 &  53.22 $\pm$ 1.38 \\
 5--10 &  42.40 $\pm$ 1.10  \\
 10--20 &  35.49 $\pm$ 0.92 \\
 20--40 &  26.89 $\pm$ 0.70 \\
 40--60 & 18.39 $\pm$ 0.48  \\
 60--80 &  10.97 $\pm$ 0.29 \\
 80--100 & 4.47 $\pm$ 0.14 \\
 \hline
\hline
 \end{tabular}
 \end{center}
\end{table} 

\subsection{Track selection and particle identification}
The charged tracks coming from the primary vertex are selected in the pseudorapidity interval $|\eta| <$ 0.8 with \pt $>$ 0.15 \GeVc. This ensures the uniform acceptance for the central barrel detectors. The high quality tracks are chosen based on selection criteria as done previously in Ref.~\cite{Adam:2016bpr}. The \kstar and \phim mesons are reconstructed 
from the charged tracks which have crossed at least 70 out of maximum 159 horizontal segments  along the transverse readout plane of the TPC.  
 The contamination from  secondary particles originating from  weak decays and beam background events are reduced by applying a
  selection on the distance of closest approach to the primary vertex in the transverse plane $(\mathrm{DCA}_{xy})$ and along the longitudinal direction $(\mathrm{DCA}_{z})$. 
  A $p_{\rm{T}}$-dependent cut of  DCA$_{xy} (\pt) <$ $(\mathrm{0.0105} + \mathrm{0.035} \pt^{-1.1})$ cm, with \pt in \GeVc, is used, which is less than 7 times its resolution. The track DCA$_{z}$ is required to be less than 2 cm~\cite{ALICE:2015ial}.
  The decay daughters (pions and kaons) of resonances are identified by measuring the specific ionization energy loss $(\dEdx)$ in the detector gas of the TPC and  their time-of-flight information using the TOF~\cite{Akindinov:2013tea}. 
  The \dEdx resolution of the TPC is denoted as  $\sigma_{\mathrm{TPC}}$ and the charged tracks are identified as pions and kaons if the mean specific energy loss measured by the TPC is within 6$\sigma_{\mathrm{TPC}}$, 3$\sigma_{\mathrm{TPC}}$, and 2$\sigma_{\mathrm{TPC}}$ from the expected $\langle\dEdx\rangle$ values in the momentum range $p$ $<$ 0.3 \GeVc, 0.3 $<$ $p$ $<$ 0.5 \GeVc and $p$ $>$ 0.5 \GeVc, respectively.
  In addition to the TPC, if the TOF information is available, then the charged tracks are identified by requiring the  time-of-flight values within  3$\sigma_{\mathrm{TOF}}$ of the expected values  for the full momentum range.

\subsection{Yield extraction}
\begin{figure}[!h]
    \begin{center}
      \includegraphics[width =0.8\textwidth]{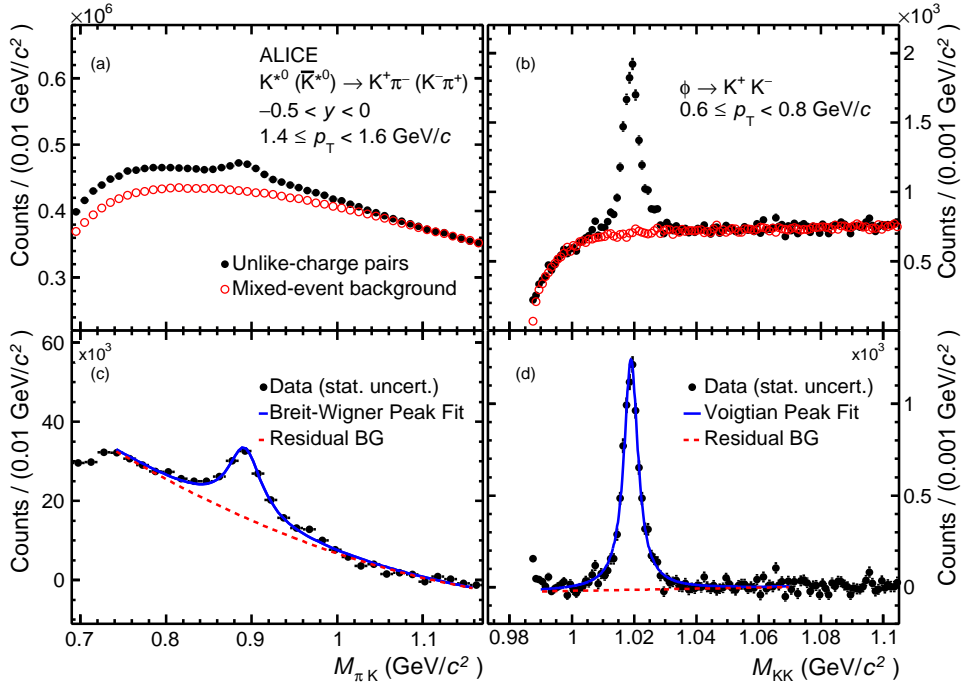}
    \end{center}
    \caption{Invariant mass distributions for \kstar and \phim in the multiplicity class 0--100$\%$ and  transverse momentum range 1.4 $\leq$ \pt  $<$ 1.6 \GeVc  and 0.6 $\leq$ \pt  $<$ 0.8 \GeVc, respectively. In the upper panels, (a) and (b), black markers show the unlike-sign invariant mass distributions and red markers show the normalized mixed event background. After the background subtraction the signals are shown in the lower panels (c) and (d). The \kstar peak is described by a Breit-Wigner function whereas the \phim peak is fitted with a Voigtian function. The residual background is described by the 2$^{nd}$ order polynomial function.}
    \label{fig:Invmass}
\end{figure}

  The \kstar and \phim resonances are reconstructed  from their decay products using the invariant mass technique.                    
 The invariant mass distributions are obtained from unlike charge  $\pi$K (for  \kstar) and KK (for \phim) pairs in the same event. The distributions exhibit a signal peak and a large combinatorial background from  the uncorrelated $\pi$K (KK) pairs. The combinatorial background is estimated using two methods, mixed-event and like-sign. In the mixed-event method, the tracks from one event are paired with oppositely charged tracks from other events. Each event is mixed with five other events to reduce the contribution of statistical uncertainty from the background distribution. The events which are mixed are selected to have similar characteristics like the longitudinal position of primary vertex (v$_{z}$) must differ by less than 1 cm and  the multiplicity percentiles computed using the V0A amplitude must differ by less than 5$\%$. The mixed-event distributions for  \kstar(\phim) are normalized in the mass region 1.1 $< m_\mathrm{inv} <$ 1.15 GeV$/c^{2}$ \mbox{(1.04 $< m_\mathrm{inv} <$ 1.15 GeV$/c^{2}$)} 
 that is approximately five $\sigma$
  away from the mass peak of each particle. 
In the like-sign method, tracks of identical charges from the same events are paired and the invariant mass distribution for the uncorrelated background is obtained as the geometric mean 2$\sqrt{n^{++}\times n^{--}}$, where $n^{++}$ and $n^{--}$ are the number of positive-positive and negative-negative pairs in each invariant mass bin, respectively. The mixed-event technique is the default method used for the extraction of yield both for  \kstar and  \phim whereas the like-sign background is used for the estimation of the systematic uncertainty. In Fig.~\ref{fig:Invmass}, panels (a) and (b) show the invariant mass distributions of ${\rm{\pi^{\mp} K^{\pm}}}$  and ${\rm{K^{+}K^{-}}}$  pairs from the same events and the mixed-events in the transverse momentum interval \mbox{1.4 $\leq$\pt  $<$ 1.6 } and \mbox{0.6 $\leq$ \pt  $<$ 0.8 \GeVc}  for 0--100$\%$ in p--Pb collisions, respectively.  
The  ${\rm{\pi^{\mp} K^{\pm}}}$ and ${\rm{K^{+}K^{-}}}$ invariant mass distributions after mixed-event background subtraction are shown in panels (c) and (d) of Fig.~\ref{fig:Invmass}, respectively, where the characteristic signal peak is observed on top of the residual background.  
The residual background arises due to correlated pairs from jets, misidentification of the decay daughters of  resonances and decay of other particles~\cite{Adam:2016bpr}. 
The raw yields of resonances are extracted in each \pt bin and multiplicity class. The signal peak is fitted with a Breit-Wigner and a Voigtian function (convolution of Breit-Wigner and Gaussian functions) for \kstar and \phim, respectively. A second order polynomial function is used to describe the shape of the residual background for both  resonances. The signal peak fit is performed in the range 0.75 $< \rm{M}_{K\pi} <$ 1.15 GeV$/c^{2}$ \mbox{(0.99 $< \rm{M}_{KK} <$ 1.07 GeV$/c^{2}$ )} for \kstar (\phim ). 
The widths of the  \kstar and  \phim are fixed to their PDG values  \mbox{$\Gamma$(\kstar)  = 47.4 $\pm$ 0.6 MeV$/c^{2}$, $\Gamma$(\phim)  = 4.26 $\pm$}
0.04 MeV$/c^{2}$~\cite{Tanabashi:2018oca}, whereas the resolution parameter of the Voigtian function for \phim is kept as a free parameter. The measured resolution of the  \phim mass as a function of \pt ($\sigma$ of Gaussian) varies between 1 and 3 MeV$/c^{2}$. The sensitivity to the choice of the fitting range, the normalization interval, the shape of the background function, the width and resolution parameters have been studied by varying the default settings, as described in Sec. 2.4.  In minimum bias collisions,  \kstar(\phim) production is measured in the \pt range from 0 to 20 \GeVc (0.4 to 16 \GeVc). With the available data samples, \kstar production is measured up to \pt  = 15 \GeVc in 0--5$\%$ and 5--10$\%$, up to 
\pt  = 20 \GeVc in 10--20$\%$, 20--40$\%$ and 40--60$\%$,  up to 10 \GeVc in 60--80$\%$ and up to 6 \GeVc in 80--100$\%$ multiplicity classes,
while the \phim production is measured up to \pt  = 16 \GeVc in 0--5$\%$, 5--10$\%$, 10--20$\%$, 20--40$\%$, \pt  = 12 \GeVc in 40--60$\%$, \pt  = 10 \GeVc in 60--80$\%$
 and \mbox{\pt  = 6 \GeVc} in 80--100$\%$ multiplicity class.  \\
The raw transverse momentum distributions are normalized by the number of accepted events and corrected for the branching ratio, detector acceptance and reconstruction efficiency ($A \times \epsilon_{\rm rec}$) and, signal loss.
The correction factor due to the vertex reconstruction efficiency is negligible in all multiplicity classes.
The $A \times \epsilon_{\rm rec}$ is obtained from the Monte Carlo simulation (MC) based on the DPMJET~\cite{Roesler:2000he}  event generator and the interaction of the generated particles passing through the ALICE detector geometry is modeled using GEANT3~\cite{Brun:1119728}. It is defined as the ratio of the reconstructed \kstar (\phim) to the generated \kstar (\phim), both in the rapidity interval $-$0.5 $<$ $y$ $<$ 0, and determined as a function of \pt.
The same track and particle identification (PID) selection criteria are applied to the decay daughter of resonances in MC as  are used in the analysis. The shape of the generated \pt distributions are different from the measured \pt distributions, therefore a re-weighting procedure is used, in which the generated distributions are weighted to match the measured distributions. The effect of the re-weighting procedure on $A \times \epsilon_{\rm rec}$ is $\approx2-5\%$ at  low \pt ($<1$~\GeVc) and negligible for \pt $>1$~GeV$/c$.
The re-weighted $A \times \epsilon_{\rm rec}$ is used to correct the raw\pt distribution. No significant multiplicity dependence of $A \times \epsilon_{\rm rec}$ is observed, therefore the raw \pt spectra in the various multiplicity classes are corrected with  the minimum bias $A \times \epsilon_{\rm rec}$ values. 
The signal loss corrections that account for the loss in \kstar and \phim yields  caused by  the event selection with minimum bias trigger, rather than all NSD events, are found to be negligible in the measured \pt range. 
The minimum bias \pt spectra are normalized to the fraction of NSD events, which is 0.992.

\subsection{Systematic uncertainties}
The sources of systematic uncertainties of the measurement of \kstar and \phim
production are  signal
extraction, track selection criteria, particle identification, global
tracking efficiency, uncertainty in the material budget of the ALICE detector and
the hadronic interaction cross section in the detector material. 
A similar approach is adopted as used for the systematic uncertainty study of  \kstar and  \phim in \pPb collisions at \snn  = 5.02 TeV~\cite{Adam:2016bpr}. No multiplicity dependence of the systematic effects is observed, therefore the systematic uncertainties of minimum bias \pt spectra are propagated for all multiplicity event classes studied.  
 A summary of systematic uncertainties for  \kstar (\phim) in two transverse momentum intervals, 0 $<$ \pt $<$ 4 \GeVc (0.4 $<$ \pt $<$ 4 \GeVc) and 4 $<$ \pt $<$ 20 \GeVc (4 $<$ \pt $<$ 16 \GeVc) are given in Table~\ref{tab_systematic}. The uncertainties due to signal extraction include  variations of the signal peak fitting range, variations of width and mass resolution, mixed--event background normalization region, choice of residual background function and combinatorial background. The fitting range of the $\pi$K (KK) invariant mass distribution is varied by $\approx$ 50 (5) MeV$/c^{2}$ on each side of the signal peak. The normalization range of the $\pi$K (KK) invariant mass distributions differed by approximately 150 (50) MeV$/c^{2}$ with  respect to the default value. The width of the resonances is fixed for the default fit whereas it is kept free for systematic studies. 
The residual background is fitted with a first--order and third--order polynomial function  for the systematic studies of the signal extraction. For \phim resonance, the effect of the variation of the resolution parameter ($\sigma$ of the Gaussian) on the yield is also included in the systematic uncertainties. The combinatorial background from the like-sign method is used for systematic studies. The contribution of systematic uncertainties due to the signal extraction is 7.5--8\% for  \kstar and 2.8--4.5\% for  \phim.   
   The systematic effects due to the charged track selection are studied by varying  the criteria based  on the number
   of crossed readout rows in the TPC and the distance of closest approach to the primary vertex of the collision~\cite{ALICE:2015ial}.
   The relative contribution of uncertainties  due to the track selection are 2--3 \% for  \kstar and about 4.4--5.5\% for the  \phim. 
For the PID systematic uncertainty, the selections based on the TPC d$E/$d$x$ and TOF time-of-flight are varied. 
Three variations are taken where one is a momentum dependent PID selection of 5$\sigma_{\mathrm{TPC}}$ (0 $<$ $p$ $<$ 0.3), 2.5$\sigma_{\mathrm{TPC}}$ (0.3 $<$ $p$ $<$ 0.5), 1.5$\sigma_{\mathrm{TPC}}$ ($p$ $>$ 0.5) with 3$\sigma_{\mathrm{TOF}}$, and two momentum-independent selection; 2$\sigma_{\mathrm{TPC}}$ with  3$\sigma_{\mathrm{TOF}}$  and 2$\sigma_{\mathrm{TPC}}$ only, for both \kstar and \phim. 
This results in systematic uncertainties of 4.3--5\% for  \kstar and 1.9--3.5\%  for the  \phim. 
 The uncertainty related to global tracking arises from the difference in the ITS-TPC track matching efficiency in data and MC.
  It is estimated from the single charged track uncertainty by taking the linear sum of the uncertainties of the two charged tracks which are used to reconstruct the resonances. It contributes to the systematic uncertainties with  2--3.2\% and 2--2.3\% for  \kstar and  \phim, respectively. The material budget systematic effects account for the uncertainties in the estimation of the ALICE detector material budget and is estimated to be 1.2\%  for  \kstar and
 2.2\% for  \phim at low \pt. It is negligible at \pt $>$ 4 \GeVc for  both \kstar and \phim.
The systematic uncertainty due to the hadronic interaction cross section in the detector material is estimated to be
1.9\% for  \kstar and 2.4\%  for $\phim $ at low \pt, and negligible for \pt $>$ 4  \GeVc. The effects of material budget and hadronic interaction are evaluated by combining the uncertainties of the two charged tracks ($\pi$, K for \kstar and two K for \phim) according to the kinematics of the decay.
The systematic uncertainties  of the material budget and the hadronic interaction cross section were taken from~\cite{Adam:2016bpr}. The total systematic uncertainty is taken as the quadratic sum of all contributions and varies as 9.6--10.2\%  for  \kstar and 6.7--8.3\% for  \phim. The sources of systematic uncertainties that are multiplicity-dependent and uncorrelated  across different multiplicity classes are also estimated. The systematic uncertainties due to signal extraction and PID are fully uncorrelated, whereas global tracking, track selection criteria, material budget and hadronic cross section are correlated among event multiplicity classes.
 \begin{table}
\begin{center}
\caption{The sources of systematic uncertainties for \kstar and \phim yields 
        in  \pPb collisions at \snn  = 8.16 TeV.   For each source, the average uncertainties are listed for the low and  high--\pt intervals.}
 \label{tab_systematic}
{
\begin{tabular}{lllll}
\hline \hline
Systematic variation & \multicolumn{1}{c} ~~~~~~~~~~~~~~~\kstar~~~~~~~~ & \multicolumn{2}{c}~~~~~~~~~~~~~~~~~~~~~~~~~~~~\phim \\                               
\hline  &
\multicolumn{4}{c}{ ~~~~~~~~~~~~~~~~~~~~~~~~~~~~~~~\pt (\GeVc)~~~~~~~~~~~~~~~~~~~~~~~~~}\\
 			
\hline  
    \multicolumn{1}{c} {}  &\multicolumn{1}{c}{0.0-4.0} &\multicolumn{1}{c}{4.0-20.0}      &\multicolumn{1}{c}{ 0.4--4.0} &\multicolumn{1}{c}{4.0--16.0} \\
\hline
Yield extraction ($\%$)       &    \multicolumn{1}{c}{7.5}     & \multicolumn{1}{c}{8.0}                 &  \multicolumn{1}{c}{2.8}             &  \multicolumn{1}{c}{4.5} \\
Track selection ($\%$)        &    \multicolumn{1}{c}{3.0}     & \multicolumn{1}{c}{2.0}                 &  \multicolumn{1}{c}{4.4}             &  \multicolumn{1}{c}{5.5}  \\
Particle identification ($\%$)&    \multicolumn{1}{c}{4.3}     & \multicolumn{1}{c}{5.0}                 &  \multicolumn{1}{c}{1.9}             &  \multicolumn{1}{c}{3.5}  \\
Global tracking efficiency ($\%$)&  \multicolumn{1}{c}{2.0}    & \multicolumn{1}{c}{3.2}                 &  \multicolumn{1}{c}{2.0}             &  \multicolumn{1}{c}{2.3}   \\
Material budget ($\%$)         &   \multicolumn{1}{c}{1.2}     &  \multicolumn{1}{c}{$\textless$ 0.5}    &  \multicolumn{1}{c}{2.2}             &  \multicolumn{1}{c}{$\textless$ 0.5} \\
Hadronic Interaction ($\%$)    &   \multicolumn{1}{c}{1.9}     &  \multicolumn{1}{c}{$\textless$ 0.5}    &  \multicolumn{1}{c}{2.4}             &  \multicolumn{1}{c}{$\textless$ 1} \\ 
\hline
Total ($\%$)                           & \multicolumn{1}{c}{9.6}  &  \multicolumn{1}{c}{10.2}      &  \multicolumn{1}{c}{6.7} &   \multicolumn{1}{c}{8.3} \\
\hline \hline
\end{tabular}
}
\end{center}
 \end{table}

\section{Results and discussion}
\subsection{Transverse momentum spectra}
\begin{figure}[tb!]
    \begin{center}
       \includegraphics[width = 0.47\textwidth]{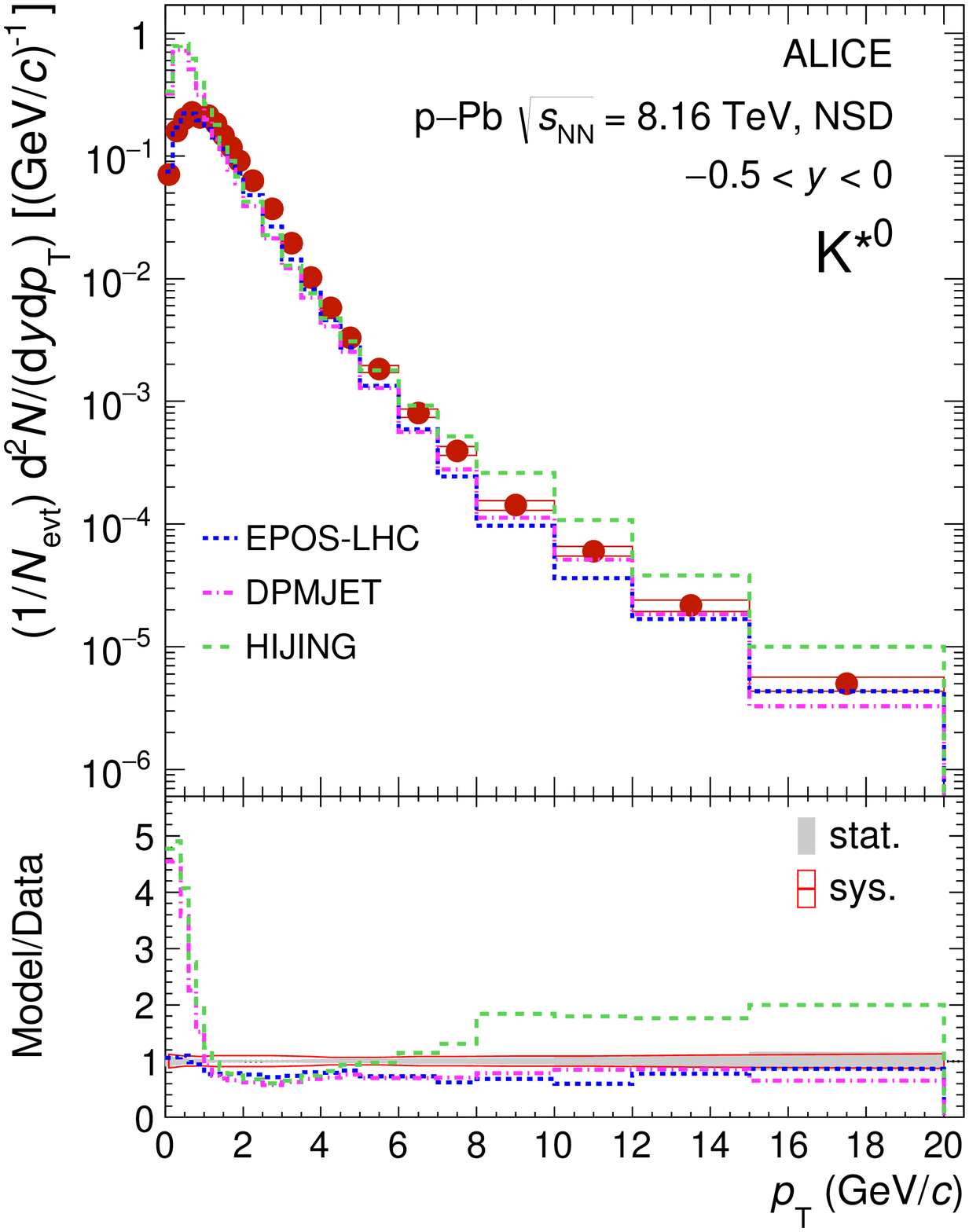}
      \includegraphics[width = 0.47\textwidth]{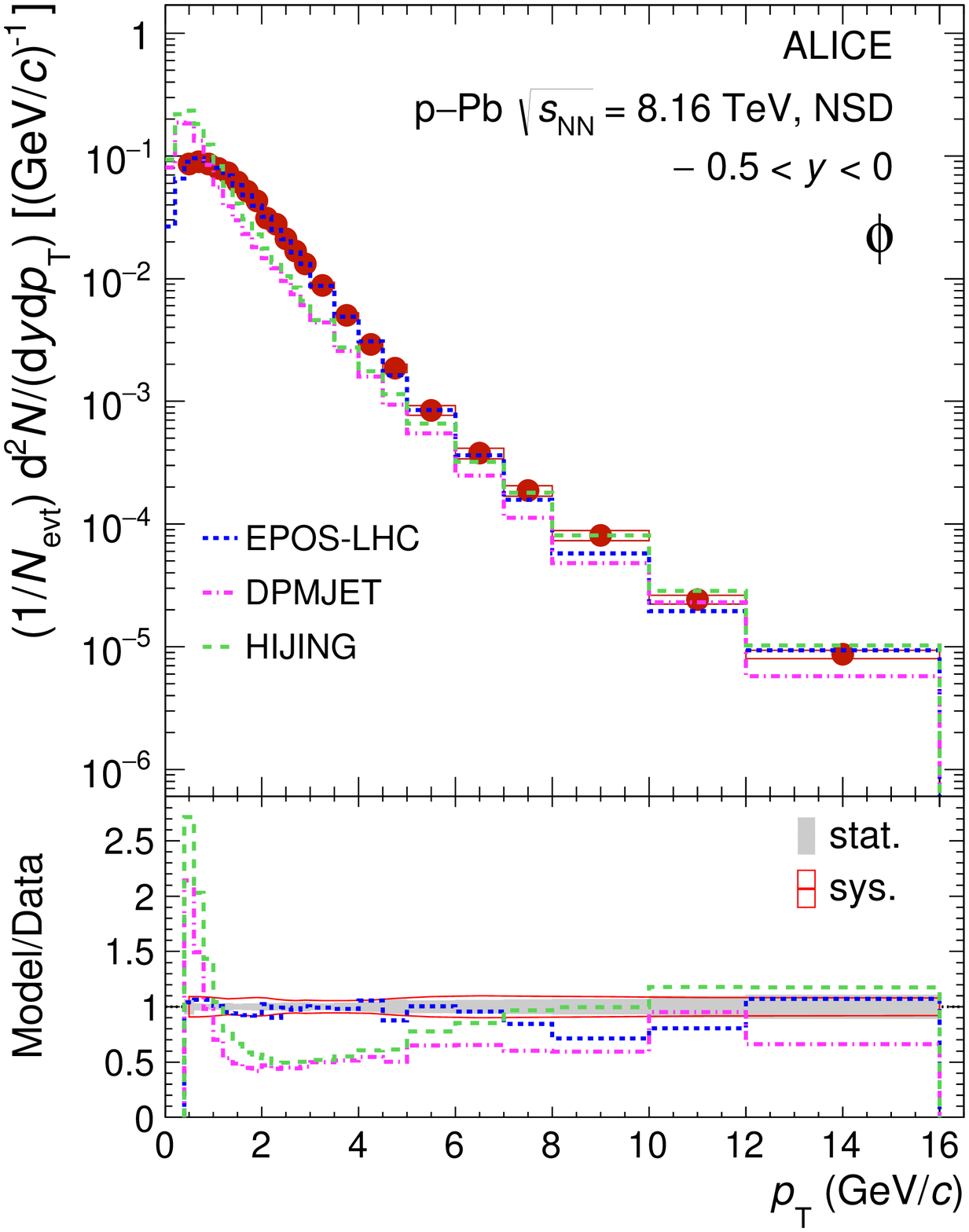}
   \end{center}
 \caption{Top panels: Transverse momentum spectrum of  \kstar (left) and \phim (right) as a function of \pt for the NSD events, measured in the rapidity interval $-$0.5 $< y$  $<$ 0 for p--Pb collisions at \snn  = 8.16 TeV. The statistical and systematic uncertainties are shown as bars and boxes, respectively. The NSD spectrum is compared with the predictions from EPOS-LHC~\cite{Pierog:2013ria,Werner:2013tya}, DPMJET~\cite{Roesler:2000he} and HIJING~\cite{Gyulassy:1994ew}. Bottom panels: The ratios of \pt spectra from model to data. The shaded bands around unity describe the statistical and systematic uncertainties of the data point.}    
    \label{fig:nsdspectra}
   \end{figure}

The measurement of \kstar(\phim) production performed in the rapidity interval $-$0.5 $<$ y $<$ 0 up to \pt  = 20 (16) \GeVc in p--Pb collisions at \snn  = 8.16 TeV is reported. Figure~\ref{fig:nsdspectra} shows  \pt spectra of \kstar (left panel) and \phim  (right panel) for NSD events. These are compared with the predictions
  from EPOS-LHC~\cite{Pierog:2013ria,Werner:2013tya}, DPMJET ~\cite{Roesler:2000he} and HIJING ~\cite{Gyulassy:1994ew} models.

  \begin{figure}[ht!]
  	\begin{center}
  		\includegraphics[width = 0.45\textwidth]{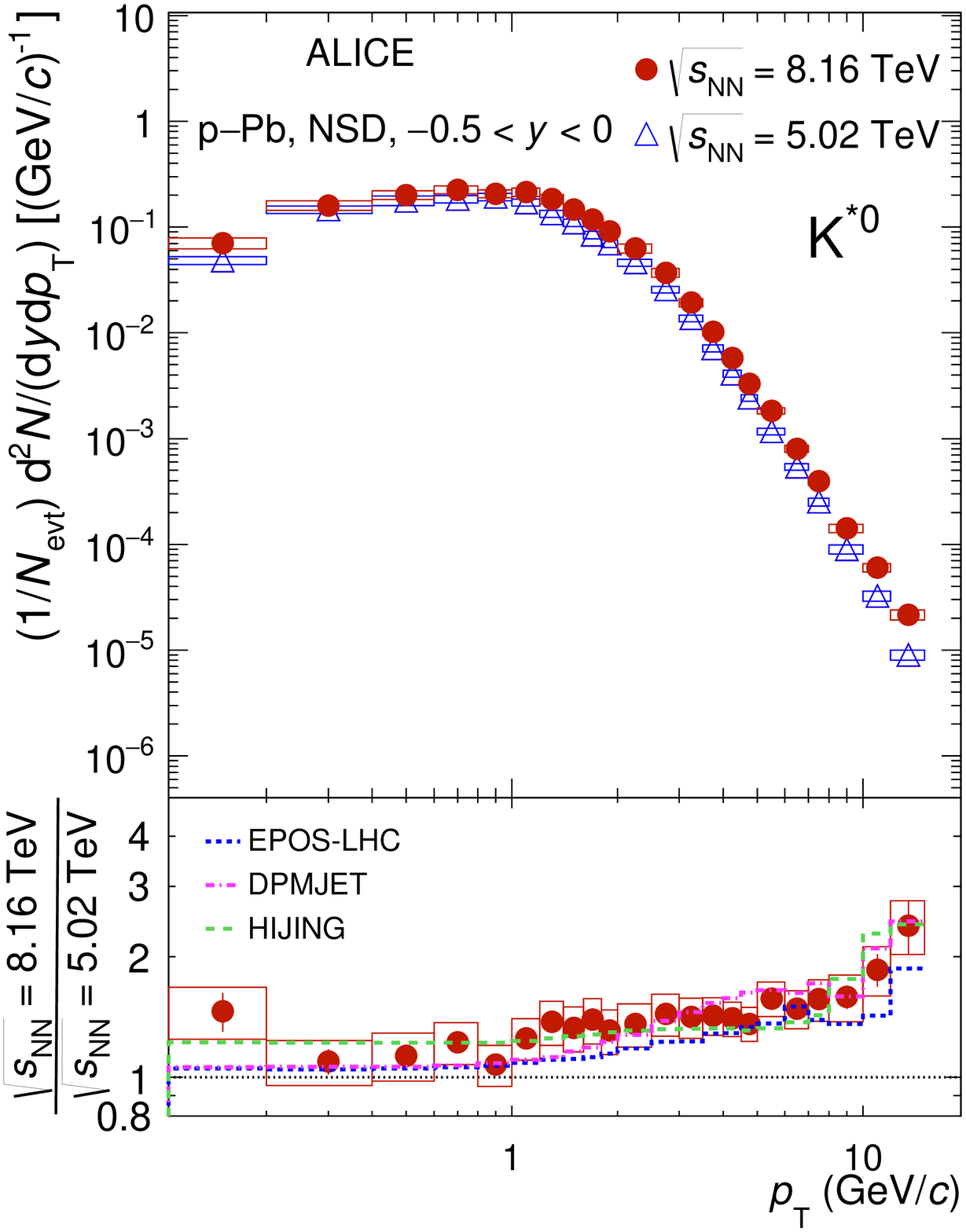} 
	         \includegraphics[width = 0.45\textwidth]{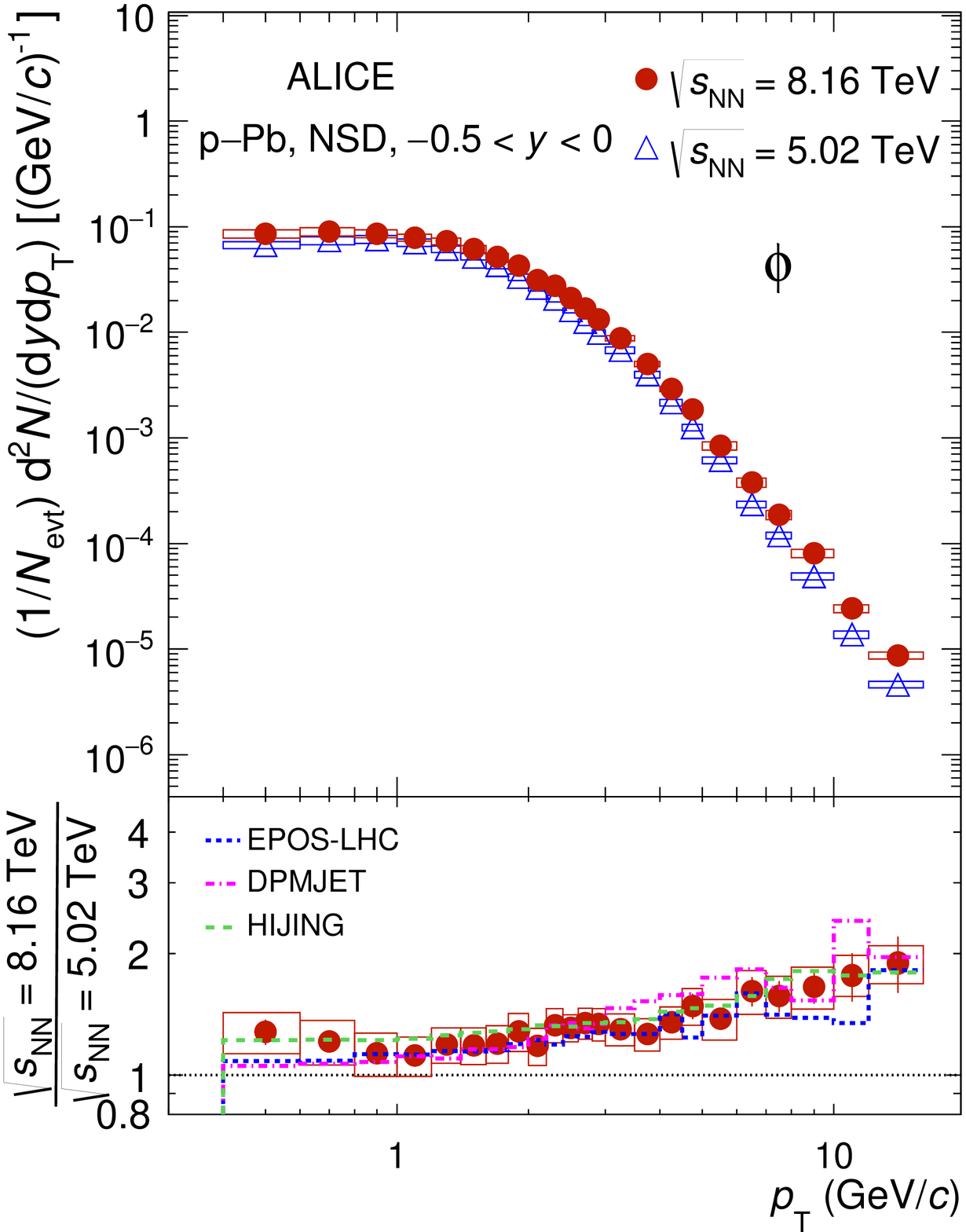}
  		\end{center} 	
		 \caption{Top panels: Energy dependence comparison of the transverse momentum spectra  of  \kstar (left) and \phim (right) as a function of\pt for the NSD events, measured in the rapidity interval $-$0.5 $<$ $y$  $<$ 0 for p--Pb collisions at \snn  = 5.02 and 8.16 TeV. Bottom panels: The ratio of \pt spectrum at   \snn  = 8.16 TeV to the \pt spectrum at \snn  = 5.02 TeV. The ratio is compared with the predictions from EPOS-LHC~\cite{Pierog:2013ria,Werner:2013tya}, DPMJET~\cite{Roesler:2000he} and HIJING~\cite{Gyulassy:1994ew}. The statistical and systematic uncertainties are shown as bars and boxes, respectively.}    
  	\label{fig:nsdenergyspectra}
\end{figure}

The bottom panels of  Fig.~\ref{fig:nsdspectra} show the ratios of \pt spectra from these models to the data.
  The EPOS Monte Carlo event generator is a hadronic interaction parton model based on Gribov's Reggeon field theory formalism which includes the feature of collective hadronization and the core-corona mechanism from pp to A--A collisions~\cite{Werner:2013yia,Drescher:2000ha,Werner:2007bf}. If the string segments of the final state parton have high energy density that region is known as the ``core'', whereas the region with strings of low energy density surrounding the core is called the ``corona''. 
The core evolves hydrodynamically and subsequently hadronizes to form the bulk of the system 
 whereas the strings in the corona region break through the production of quark-antiquark pairs, which hadronize as fragmentation processes in vacuum.
EPOS-LHC~\cite{Pierog:2013ria} is a tune of EPOS1.99~\cite{Pierog:2009zt} 
that incorporates a parameterization of flow based on  LHC data.
The EPOS1.99 model is different from EPOS2.x~\cite{Werner:2010aa} and EPOS3.x~\cite{Werner:2013tya} as it does not use the complete 3D hydro calculation followed by the hadronic cascade but instead relies on the fast covariant approach.
It  describes various observables in minimum bias heavy-ion collisions as well as small collision systems up to a few \GeVc at LHC energies.
DPMJET is a QCD-inspired dual parton model based on the Gribov-Glauber approach that treats the soft and hard scattering interaction processes differently. HIJING combines the perturbative QCD process with soft excitation, the production of multiple minijets,  the interactions of jets in dense hadronic matter, and nuclear shadowing of parton distribution functions.
For the \kstar resonance, at low \pt ( $<$ 1  \GeVc), DPMJET and HIJING models overestimate the data, whereas EPOS-LHC model gives a good description of the \pt spectrum. At  \pt  $>$ 1 \GeVc, DPMJET and EPOS-LHC underestimate and closer to the data, however HIJING model underestimates (similar to the DPMJET and EPOS-LHC)  for  1 $<$ \pt  $<$ 5 \GeVc and overestimates  for \pt  $>$ 5 \GeVc.
The EPOS-LHC model describes the \phim \pt spectrum relatively better than the DPMJET and HIJING for all \pt. However, HIJING models gives a good description of \pt distributon of \phim resonance for \pt $>$ 6 \GeVc. 
The EPOS-LHC model, where a different parametrization of flow is introduced in small collision systems like pp than the large volume produced in heavy-ion collisions, gives a better description of the transverse momentum distributions for both \kstar and \phim in p--Pb collisions.
Figure~\ref{fig:nsdenergyspectra} shows  the \snn dependence of the transverse momentum spectra of \kstar and \phim for NSD events in \pPb collisions. The upper panels of Fig.~\ref{fig:nsdenergyspectra} show a comparison of the transverse momentum spectra of \kstar and \phim at \snn  = 5.02 and 8.16 TeV whereas  the lower panels show the ratio of
the  \pt- differential yield  at \mbox{\snn  = 8.16 to 5.02 TeV} and its comparison with the results obtained from
models~\cite{Werner:2013tya,Pierog:2013ria,Roesler:2000he,Gyulassy:1994ew}.
The uncertainties of the ratios are  obtained as the sum in quadrature of the uncertainties of the spectra at the two energies, which  are largely uncorrelated.
Up to \pt 1 \GeVc, the differential yield ratio seems to independent of \pt and collision energy. The values are consistent with unity within uncertainties.
It suggests that the particle production in the soft scattering region is not strongly dependent  on collision energy. The differential yield ratios increases as a function of  \pt for \pt $\gtrsim$ 1 \GeVc. 
Similar behavior is also observed in pp collisions in Ref.~\cite{Acharya:2019wyb}. 
The \pt differential yield ratios obtained from EPOS-LHC, DPMJET, and  HIJING are consistent with the measurements within the systematic uncertainties and  reproduce well the energy dependence trend  for \kstar and \phim in \pPb collisions.

\begin{figure}[tb!]
    \begin{center}
      \includegraphics[width = 0.45\textwidth]{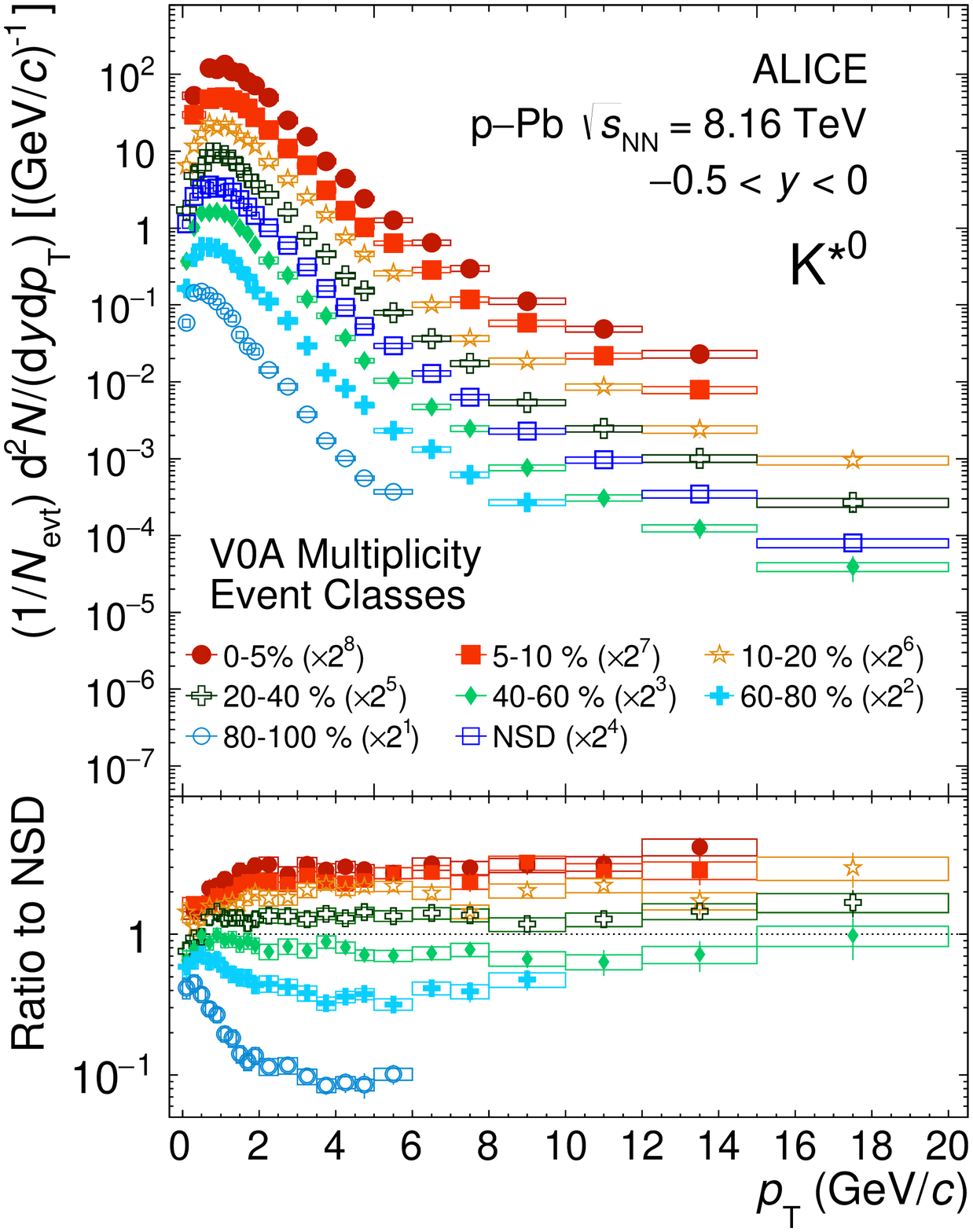}
   \includegraphics[width = 0.45\textwidth]{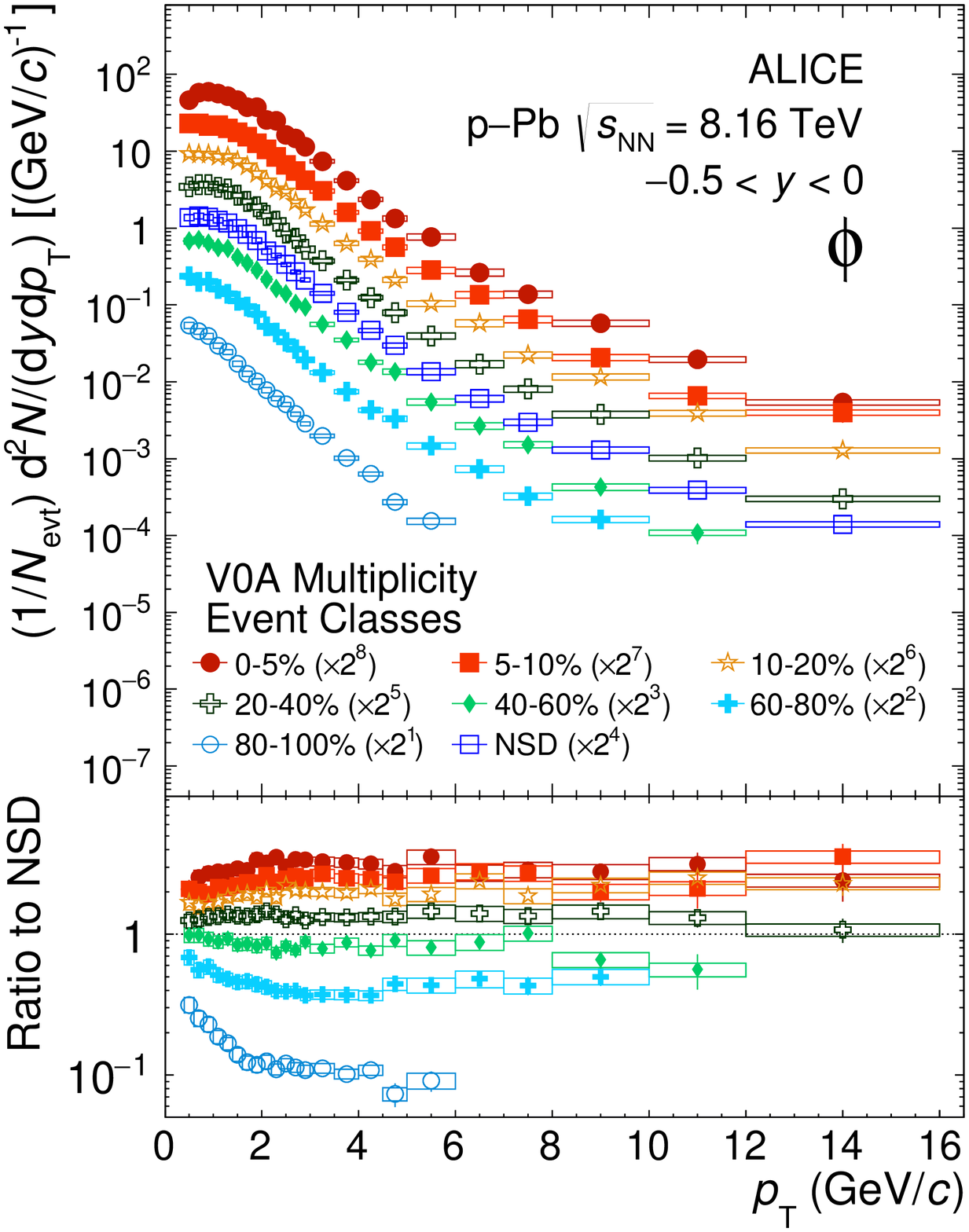}
 \end{center}
  \caption{Top panels: The transverse momentum spectra of  \kstar (left) and \phim (right) for various multiplicity classes, measured in the rapidity interval $-0.5 <$ $y$  $<$ 0 for p--Pb collisions at \snn  = 8.16 TeV. Bottom panels: The ratios of \pt spectra of given event multiplicity classes to the NSD spectra are shown. The statistical and systematic uncertainties are shown as bars and boxes, respectively.}
    \label{fig:ptspectra}
     \end{figure}

Figure~\ref{fig:ptspectra}  shows the transverse momentum distributions of \kstar (left panel) and \phim (right panel) in various multiplicity classes. The ratios of \pt spectra in various multiplicity classes to the \pt spectrum for NSD events are shown in the bottom panels of~Fig.~\ref{fig:ptspectra}.

For \pt $\lesssim$ 4 \GeVc, the slopes of the \pt spectra increase from low to high multiplicity classes, whereas the spectral shapes  are similar at high \pt for all multiplicity classes. This indicates that processes like radial flow, which lead to a change in the shape of the \pt spectra for various multiplicity classes, dominate mainly at  low \pt~\cite{Acharya:2019yoi}. The increase in the slope of \pt spectrum with multiplicity is reflected in Fig.~\ref{fig:scaleddndymeanpt} for \meanpt as a function of multiplicity.
A similar behavior was also observed for \kstar and \phim in p--Pb collisions~\cite{Adam:2016bpr} at \snn = 5.02 TeV. The hardening of the \pt spectra with charged particle multiplicity was also reported for inclusive charged hadron spectra, $\pi$, K, p, K$^{0}_{S}$, $\Lambda$, $\Xi$, and $\Omega$ in pp collisions at LHC energies~\cite{Acharya:2018orn,Acharya:2019bli,Acharya:2019kyh,Acharya:2019mzb}, where different models with multi-parton interactions were shown to describe these effects.
  
  \subsection{Integrated particle yield and mean transverse momentum }
\begin{figure}[p]
	\begin{center}
\includegraphics[scale=0.8]{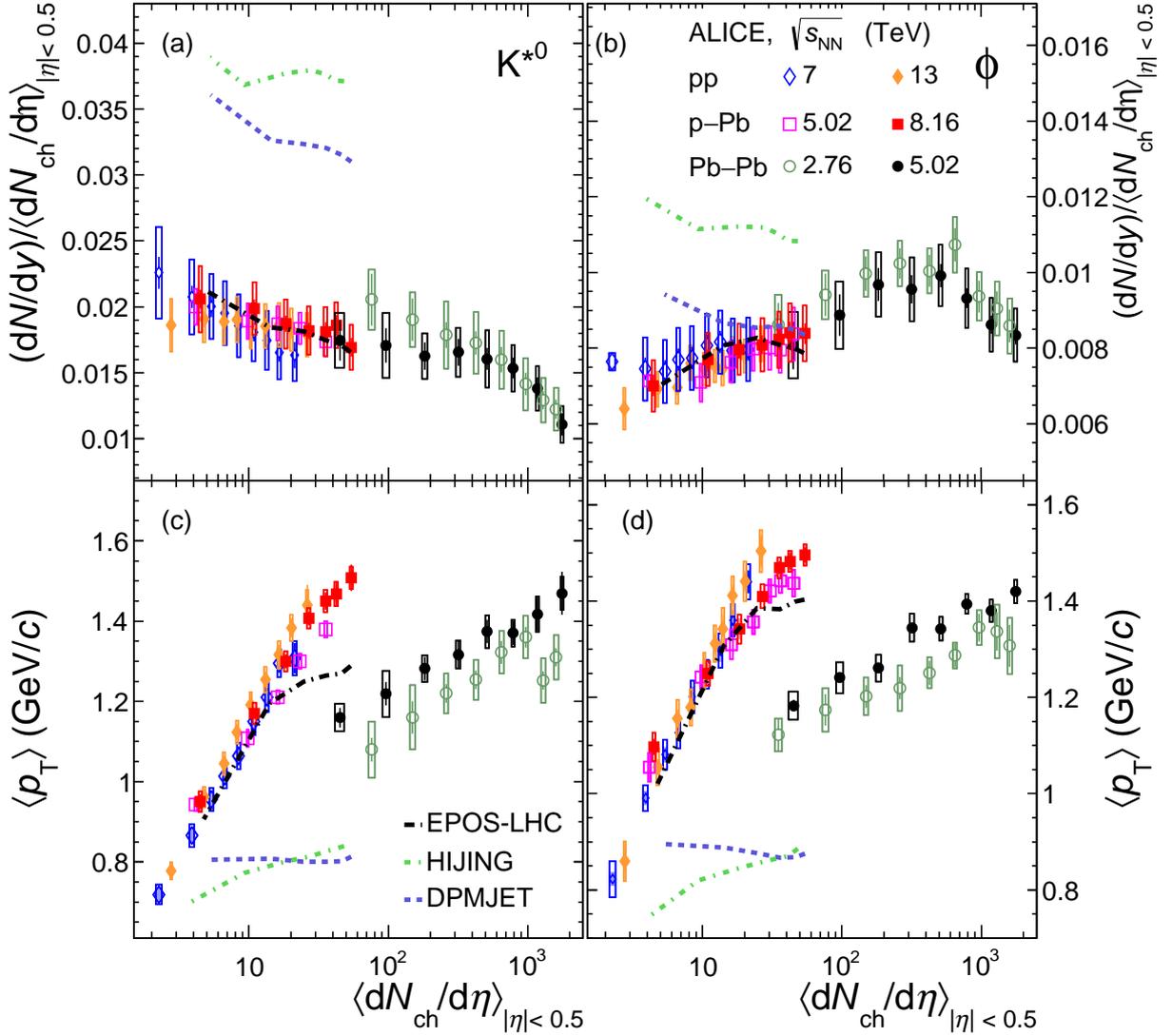}          
	\end{center}
   \caption{The multiplicity-scaled integrated yield $((\dNdyy)/(\langle\mathrm{d}N_{\mathrm{ch}}/\mathrm{d}\eta\rangle_{|\eta|<0.5}))$ (upper panels) and mean transverse momentum $(\langle p_{\mathrm{T}}\rangle)$ (bottom panels) for \kstar (left panels) and \phim (right panels)  as a function of $\langle\mathrm{d}N_{\mathrm{ch}}/\mathrm{d}\eta\rangle_{|\eta|<0.5}$ measured in the ALICE central barrel in pp collisions at $\sqrt{s}$  = 7, 13 TeV, in \pPb collisions at  \snn  = 5.02 , 8.16 TeV and  \PbPb  collisions at \snn  = 2.76 , 5.02 TeV. Measurements are compared with the predictions from EPOS-LHC~\cite{Pierog:2013ria,Werner:2013tya}, DPMJET~\cite{Roesler:2000he} and HIJING~\cite{Gyulassy:1994ew} for p--Pb collisions at \snn  = 8.16 TeV. Statistical uncertainties are represented as bars, boxes indicate total systematic uncertainties. }
  	\label{fig:scaleddndymeanpt}
 \end{figure}

  The \pt-integrated yields and mean transverse momentum are extracted from transverse momentum spectra in the measured range and using the fit function in the unmeasured region. The \phim yield is extrapolated in the unmeasured region (\pt $<$ 0.4 \GeVc) by fitting a L\'{e}vy-Tsallis functions~\cite{Tsallis:1987eu} to the measured \pt spectra in all multiplicity classes. The difference in the yield contribution  at low \pt due to different fitting functions (i.e exponential, Boltzmann, $\mathrm{m}_{T}$-exponential, Bose-Einstein and Boltzmann-Gibbs Blast-Wave function in Ref.~\cite{ALICE:2020jsh}) from the L\'{e}vy-Tsallis function is included in the systematic uncertainties. The  low--\pt extrapolation  accounts for 8.9\% (14.1\%) of the total yield 
  in the 0--5\% (80--100\%) multiplicity class. The  \kstar spectra are measured from \pt  = 0, so  low--\pt extrapolation is not needed. The contribution of the extrapolated fraction of the yield is negligible for \pt $>$ 20 \GeVc (16 \GeVc) for \kstar (\phim). The values of \dNdyy and \meanpt  of \kstar and \phim for various multiplicity classes are summarized in the Table~\ref{tab_dndy}.
  \begin{table}
 	\begin{center}
          \caption{The  values of \dNdyy  and \meanpt are presented for different multiplicity classes in p--Pb collisions at \snn $=$ 8.16 TeV. In each entry, the first uncertainty is statistical and the second is systematic. The value given in the parentheses corresponds to uncorrelated  part of the systematic
uncertainty. The fraction of total yield obtained by extrapolation (``extr.'') are also reported.}
 	\label{tab_dndy}
 		\scalebox{0.9}{
 			\begin{tabular}{cccc}
 				\hline  
 				& &  \kstar \\
 				\hline
 				Multiplicity ($\%$) & extr.   & \dNdyy   & \meanpt (\GeVc)\\
 				\hline
				0--5 & - & 0.913 $\pm$ 0.030 $\pm$ 0.086  (0.047) &1.509 $\pm$ 0.033 $\pm$ 0.028  (0.018) \\
 				5--10 & - &0.783$\pm$ 0.025 $\pm$ 0.074  (0.050)  & 1.461 $\pm$ 0.029 $\pm$0.030 (0.021)\\
 				10--20 & - & 0.644 $\pm$ 0.015 $\pm$0.060 (0.047) & 1.460 $\pm$ 0.021 $\pm$0.028 (0.020) \\
 				20--40 & - & 0.489 $\pm$ 0.009 $\pm$ 0.045 (0.028)  & 1.407 $\pm$ 0.016 $\pm$ 0.025 (0.017)\\
 				40--60 & - & 0.344 $\pm$ 0.006 $\pm$ 0.032 (0.018)  &  1.301 $\pm$ 0.014 $\pm$ 0.025 (0.014)\\
 				60--80 & - & 0.220  $\pm$ 0.004 $\pm$ 0.020 (0.016)  & 1.176 $\pm$ 0.012 $\pm$0.026 (0.020)\\
 				80--100 & - & 0.092 $\pm$ 0.002 $\pm$ 0.008 (0.006) &  0.950 $\pm$ 0.011 $\pm$ 0.026 (0.018)\\
 				NSD & - & 0.396 $\pm$ 0.004 $\pm$ 0.037 &  1.335 $\pm$ 0.008 $\pm$ 0.027\\				
 		          	\hline 
				& & \phim \\
 				\hline
 				Multiplicity ($\%$)     &   extr.  & \dNdyy   & \meanpt (\GeVc)\\
 				\hline

                                  0--5 & 0.089 &    0.455 $\pm$  0.008 $\pm$0.041 (0.026) &  1.496 $\pm$ 0.015 $\pm$ 0.022 (0.021) \\
 				5--10 & 0.088 & 0.356 $\pm$ 0.007 $\pm$ 0.033 (0.028) &  1.482 $\pm$ 0.017 $\pm$ 0.022 (0.021) \\
 				10--20 & 0.093 & 0.292 $\pm$ 0.004 $\pm$ 0.028 (0.018) & 1.468 $\pm$ 0.013 $\pm$ 0.021 (0.016)   \\
 				20--40 & 0.101 & 0.217 $\pm$ 0.003 $\pm$0.020 (0.008)  &  1.409 $\pm$ 0.011 $\pm$ 0.025 (0.010)  \\
 				40--60 & 0.104 & 0.146 $\pm$ 0.002 $\pm$ 0.014 (0.007) &  1.342 $\pm$ 0.011 $\pm$ 0.029 (0.021) \\
 				60--80 & 0.122 & 0.084 $\pm$ 0.001 $\pm$0.008 (0.004) &  1.249 $\pm$ 0.013 $\pm$ 0.025 (0.020) \\
				80--100 & 0.141 & 0.0313 $\pm$ 0.0008 $\pm$ 0.003 (0.002) &  1.097 $\pm$ 0.016 $\pm$ 0.029 (0.008) \\
 				NSD & 0.103 &  0.161 $\pm$ 0.002 $\pm$ 0.015 &  1.393 $\pm$ 0.008 $\pm$ 0.024 \\
 				\hline
 	            \end{tabular}
                 }
 		\end{center}
		\end{table}
  The multiplicity-scaled integrated yields $((\dNdyy)/(\langle \mathrm{d}N_{\mathrm{ch}}/\mathrm{d}\eta\rangle_{|\eta|<0.5}))$ for \kstar and \phim are shown in the upper panels of Fig.~\ref{fig:scaleddndymeanpt} as a function of $\langle \mathrm{d}N_{\mathrm{ch}}/\mathrm{d}\eta\rangle_{|\eta|<0.5}$. These results are compared with other ALICE measurements in pp collisions at $\sqrt{s}$  $=$ 7 and 13 TeV~\cite{Acharya:2018orn,Acharya:2019bli}, in p--Pb collisions at \snn  $=$ 5.02 TeV~\cite{Adam:2016bpr}, and in Pb--Pb collisions at  \snn $=$ 2.76 and 5.02 TeV~\cite{Adam:2017zbf, Acharya:2019qge,ALICE:2021ptz}.
The scaled integrated yields evolve smoothly as a function of multiplicity from pp, p--Pb to Pb--Pb collisions. For similar $\langle \mathrm{d}N_{\mathrm{ch}}/\mathrm{d}\eta\rangle_{|\eta|<0.5}$, these values are consistent within uncertainties for different colliding systems and at various LHC energies. This indicates that event multiplicity drives the resonance production, irrespective of the colliding systems and energies ~\cite{Adam:2016bpr, Acharya:2018orn,Acharya:2019bli}. \\
The scaled integrated yields of  \phim show a slight increase  with multiplicity from pp collisions to mid-central \PbPb collisions. The total increase is 12\% with a 1.5$\sigma$ significance between the lowest multiplicity bin and the highest multiplicity bin in \pPb collisions at \snn  = 8.16 TeV. 
Similarly scaled integrated yields of  \kstar show a slight decrease with multiplicity for all three collision systems and the total decrease is 12\% with a 1.8$\sigma$ significance for \pPb collisions at \snn  = 8.16 TeV. 
The significance  is calculated using statistical and multiplicity uncorrelated systematic uncertainties, added in quadrature.
The integrated yield ratios of resonances relative to those of longer lived particles, $\pi$, K, and p are computed to study their production mechanism.  The $\kstar/{K}$ ($\phim/\pi$)  ratio measured in p--Pb collisions at \snn = 5.02 TeV~\cite{Adam:2016bpr}
 shows a decreasing (increasing) trend  going from the lowest multiplicity to the highest multiplicity bin with a significance of 2.6$\sigma$ (1.5$\sigma$) which is discussed in the context of a hint of  a re-scattering (strangeness enhancement) effect. Future measurements of $\pi$ and K yields in \pPb collisions at \snn  = 8.16 TeV will be useful to study these effects at higher centre-of-mass energy and up to larger multiplicity.
 The model comparison with the p--Pb data shows that EPOS-LHC describes the scaled integrated yields for both \kstar and \phim 
whereas HIJING overestimates the data for all multiplicities. The DPMJET model describes the scaled integrated yield of \phim at higher multiplicities but overestimates the  \kstar at all multiplicities. 
  The \meanpt exhibits an increasing trend as a function of $\langle\mathrm{d}N_{\mathrm{ch}}/\mathrm{d}\eta\rangle_{|\eta|<0.5}$  for \kstar and \phim in various colliding systems and energies as shown in the bottom panels of Fig~\ref{fig:scaleddndymeanpt}. 
The increase in \meanpt is faster for pp and p--Pb than \PbPb and for a common multiplicity coverage the values of \meanpt in pp and \pPb are larger than \PbPb. At similar multiplicity ($\langle \mathrm{d}N_{\mathrm{ch}}/\mathrm{d}\eta\rangle_{|\eta|<0.5} \approx $ 40), the difference in \meanpt values among \PbPb, \pPb and pp collisions indicate that the geometry  and dynamics of the collision systems are different, while the scaled integrated yields of \kstar and \phim are similar for all colliding systems and energies. This indicates that the high multiplicity event sample in small collision systems has a dominantly large fraction of harder events. 

Similar studies are reported in Refs.~\cite{Abelev:2013bla,Adam:2016bpr}, where the moderate increase of \meanpt in \PbPb collisions was related to collective flow. The strong increase of $\langle p_{\mathrm{T}}\rangle$  with $\langle\mathrm{d}N_{\mathrm{ch}}/\mathrm{d}\eta\rangle_{|\eta|<0.5}$  in small collision systems can be further investigated by systematic studies of $\langle p_{\mathrm{T}}\rangle$ from different models in pp  and \pPb collisions that  incorporate processes like color reconnection, between strings produced in multi-parton interactions, different string fragmentation processes and  the core-corona mechanism. It was observed in Ref.~\cite{Acharya:2019bli} that the PYTHIA8 model with color reconnection, which introduces a flow--like effect, and the EPOS-LHC model, which uses parameterized flow, are able to reproduce the increasing trend of \meanpt as a function of multiplicity for \kstar and \phim in pp collisions at $\sqrt{s}$  = 13 TeV. The \pPb measurements are important, as in Ref.~\cite{Abelev:2013bla} it is shown that the 
 \meanpt of charged hadrons as a function of multiplicity shows a similar behavior as in pp collisions at low multiplicity whereas it seems to approach a similar but less prominent trend of saturation as in \PbPb collisions at high multiplicity.

The model comparison with p--Pb data shows that EPOS-LHC describes the increasing trend of $\langle p_{\mathrm{T}}\rangle$ with multiplicity for both \kstar and \phim, and it gives a better agreement for \phim to high multiplicity values. DPMJET and HIJING models fail to describe the observed trend in \meanpt for both \kstar and \phim and underpredict the data for all multiplicities.

\subsection{$x_{\rm{T}}$ scaling}
Particle invariant production cross sections are known to follow a scaling in the measurement of the transverse momentum spectrum for different collision energies at high \pt  using the scaling variable \mbox{$x_{\rm T}$  $=$ 2\pt$/\sqrt{s}$}~\cite{Brodsky:2005fza,Arleo:2009ch}.  
The $x_{\rm{T}}$ scaling was tested in pp collisions for identified hadrons in STAR~\cite{Adams:2006nd}, ALICE~\cite{ALICE:2020jsh} and for (non-identified) charged particles in CDF~\cite{Aaltonen:2009ne,CDF:2001hmt,Abe:1988yu}, UA1~\cite{Albajar:1989an}, and CMS~\cite{Chatrchyan:2011av}. In this paper, the validity of empirical $x_{\rm{T}}$ scaling is tested using the \kstar and \phim measurements in p--Pb collisions at \snn  = 8.16 TeV reported here and those obtained at \snn  = 5.02 TeV~\cite{Adam:2016bpr}.
	The invariant cross sections are determined from the measured invariant yield as $E\mathrm{d}^{3}\sigma/\mathrm{d}p^{3}$  = $\sigma_{\mathrm{inel}} \times {E}\mathrm{d}^{3}N/\mathrm{d}p^{3}$, where $\sigma_{\mathrm{inel}} =$ (72.5 $\pm$ 0.5) mb and (67.6 $\pm$ 0.6) mb is the inelastic cross section in pp collisions at $\sqrt{s}$  =  8.16 TeV and 5.02 TeV, respectively~\cite{Loizides:2017ack}. 

   At fixed $x_{\rm{T}}$, the invariant cross section ${E}\mathrm{d}^{3}\sigma / \mathrm{d}p^{3}$ scales as $p_{\mathrm T}^{-n}$, where the exponent of scaling $n$ depends on $x_{\rm{T}}$ and \snn, and is calculated using the following equation    
	\begin{equation}
	n(\it{x}_{\rm{T}}, \sqrt{s_{\rm{NN1}}}, \sqrt{s_{\rm{NN2}}}) = \frac {{\rm ln}(\sigma^{\rm{inv}} (\it{x}_{\rm{T}}, \sqrt{s_{\rm{NN2}}})/ \sigma^{\rm{inv}} (\it{x}_{\rm{T}}, \sqrt{s_{\rm{NN1}}}) )} {{\rm ln} (\sqrt{s_{\rm{NN1}}}/ \sqrt{s_{\rm{NN2}}} )}, 
	\label{n_xt}
	\end{equation}

  \begin{figure}[H]
    \begin{center}
    \includegraphics[scale = 0.5]{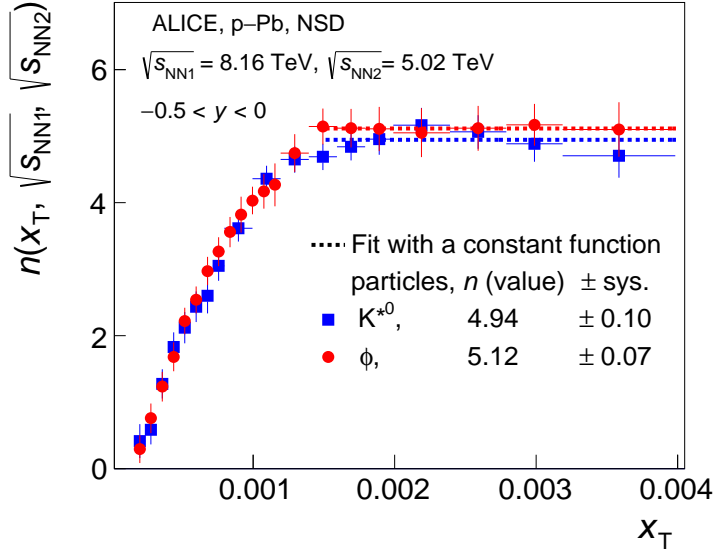}
    \end{center}
    \caption{The values of $n$ as a function of $x_{\rm{T}}$ for \kstar and  \phim in \pPb collisions.}
    \label{fig:nvsXt}
  \end{figure}
\begin{figure}[H]
    \begin{center}
    \includegraphics[scale = 0.8]{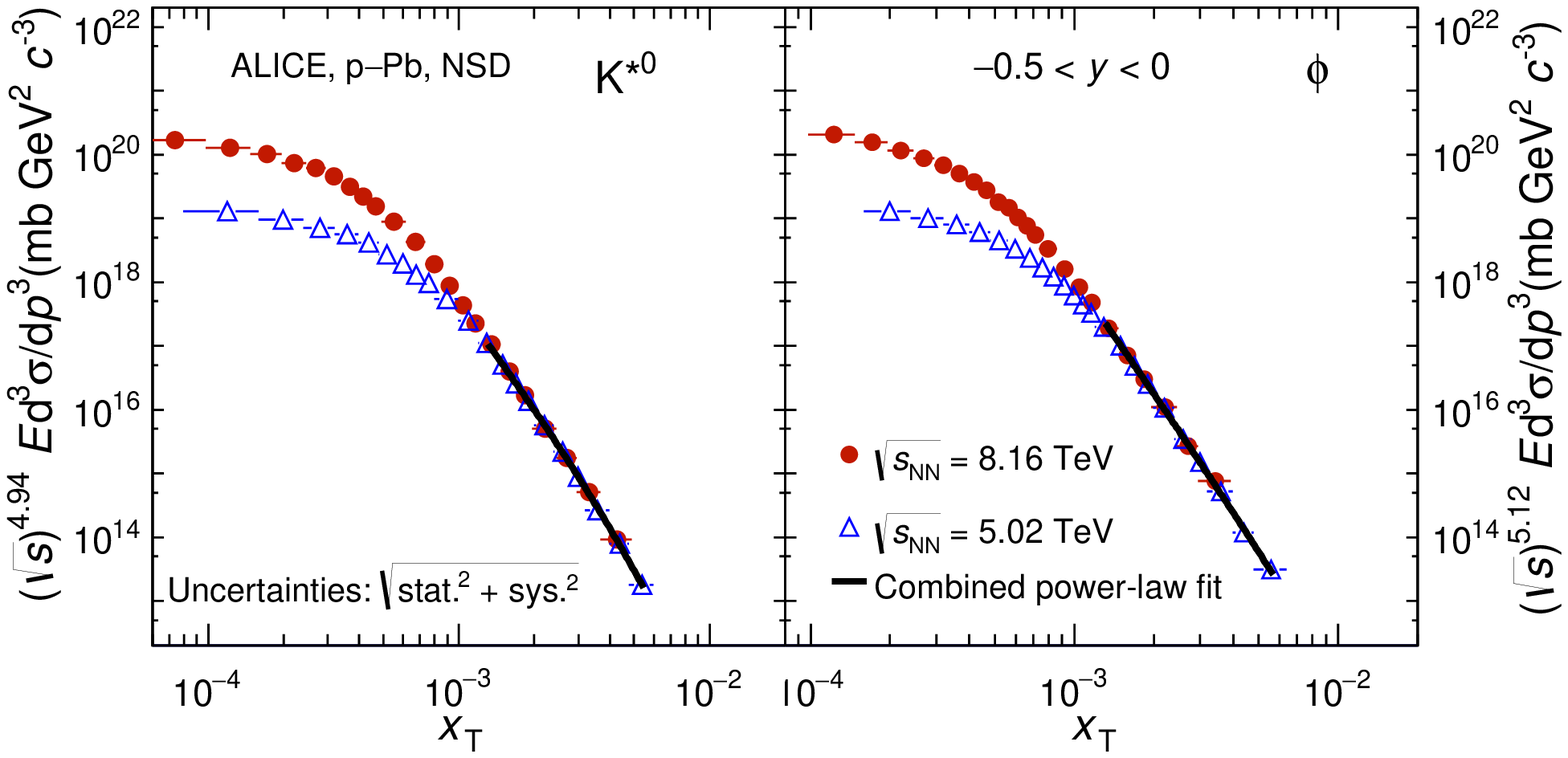}
    \end{center}
    \caption{Scaled invariant yield of \kstar and  \phim as a function of $x_{\rm{T}}$  = 2$p_{\mathrm{T}}/\snn$ in p--Pb collisions 
    at different energies \snn  = 5.02 and 8.16 TeV.}
 \label{fig:Xt}
\end{figure}

   where $x_{\rm{T}}$ = 2$\pt/\sqrt{s_{\rm{NN}}}$. The distributions of $n$ values as a function of $x_{\rm{T}}$ for \kstar and \phim are shown in Fig.~\ref{fig:nvsXt}. In the low $x_{\rm{T}}$ region, where the particle production is dominated by soft processes, the values of $n$ are found to increase with $x_{\rm{T}}$ whereas the $n$ values seem to saturate at high $x_{\rm{T}}$. The $n$ values are obtained by fitting the $n(x_{\rm{T}}, \sqrt{s_{\rm{NN}}})$ distribution by a constant function in the $x_{\rm{T}}$ range  1.3 $\times$ 10$^{-3}  <$  $\it{x}_{\rm{T}} <$ 4 $\times$ 10$^{-3}$ for both  \kstar and \phim. The $x_{\rm{T}}$ spectra for both particles are scaled by the corresponding ($\sqrt{s_{\rm{NN}}}/$GeV)$^{n}$. The best scaling is obtained in the quoted fitting range with an exponent of \mbox{$n$ = 4.94 $\pm$ 0.10 (sys.)} for \kstar and \mbox{$n$ = 5.12  $\pm$ 0.07 (sys.)} for \phim.  The systematic uncertainties on the exponent $n$ are calculated by changing the fit range in $n(x_{\rm{T}}, \sqrt{s_{\rm{NN}}})$ versus $\it{x}_{\rm{T}}$  distribution. The maximum deviation of $n$ value with respect to the default one is taken as systematic uncertainties. The $n$ values for \kstar and \phim are consistent within the uncertainties, which suggests that the ratios of particle spectra attain similar values in \pPb collisions at LHC energies. The $x_{\rm{T}}$-scaled spectra for \kstar  (left panel) and \phim (right panel) in p--Pb collisions at \snn= 5.02 and 8.16 TeV are shown in Fig.~\ref{fig:Xt}.
  
   These measurements suggest that the \kstar and \phim yields in \pPb collisions at LHC energies follow $x_{\rm{T}}$ scaling for $x_{\rm{T}} \gtrsim 10^{-3}$. Similar studies were performed in pp collisions at LHC energies for identified hadrons ($\pi^{\pm}$, K$^{\pm}$, p ($\bar{\rm{p}}$) and \kstar) with ALICE~\cite{ALICE:2020jsh} and for charged hadrons with CMS~\cite{Chatrchyan:2011av}. The $n$ values obtained  in pp collisions for all hadron species except the proton are comparable to those reported here for \kstar and \phim in p--Pb collisions. In Ref.~\cite{ALICE:2020jsh}, the proton takes a larger value of the exponent $n$, which was discussed in the context of the decrease of the baryon-to-meson ratio with increasing \pt in contrast to the nearly constant behavior of meson-to-meson ratios.  The $n$ value obtained at LHC energies is also observed to be lower than at RHIC energies, which suggests an increasing contribution of hard processes at higher centre-of-mass energies. 

A combined fit to the scaled differential cross sections of \kstar and \phim is performed with a power-law function of the form $a \times \it{x}_{\rm{T}}^{b} \times$(1+$\it{x}_{\rm{T}})^{c}$ to verify the quality of the scaling behaviour. Here, $a$, $b$, and $c$ are free parameters. The fitting is done in the region above $\it{x}_{\rm{T}}$ $\gtrsim$ 1.3 $\times$ $10^{-3}$ (shown as black curve in  Fig.~\ref{fig:Xt}), where the $\it{x}_{\rm{T}}$ scaling is observed. The $\chi^{2}/$ndf value for \kstar(\phim) is 0.16 (0.6), which confirms the good quality of the fit. In the fitting region, the measurements agree with the combined power law fits within $\approx$ 20\% for both \kstar and \phim. The measurements at \snn  = 8.16 TeV are consistent, over the accessible $x_{\rm{T}}$ range 1.3 $\times$ 10$^{-3}  <$  $\it{x}_{\rm{T}} <$ 3 $\times$ 10$^{-3}$, with empirical $x_{\rm{T}}$ scaling and with measurements from p--Pb collisions at \snn  =  5.02 TeV. This further helps  understanding and distinguishing the contributions of the soft and hard processes to particle production.
\subsection{Nuclear modification factor (\RpPb)}
In order to understand the nuclear effects, the nuclear modification factor $({R}_\mathrm{pPb})$ is an important observable in \pPb collisions.
It is calculated as : 
\begin{equation}
\it{R}_{\mathrm{pPb}} (\it{p}_\mathrm{T}) = \frac{\mathrm{d}^{2}N_{\rm{pPb}} / \mathrm{d}p_\mathrm{T}\mathrm{d} y } 
{\langle T_\mathrm{pPb} \rangle \mathrm{d}^{2}\sigma^{\mathrm{INEL}}_{\rm{pp}}/ \mathrm{d}p_\mathrm{T}\mathrm{d}y },
\end{equation}
where $\mathrm{d}^{2}N_{\rm{pPb}} / \mathrm{d}p_\mathrm{T}\mathrm{d} y$ is the  yield in \pPb collisions and $\mathrm{d}^{2}\sigma^{\mathrm{INEL}}_{\rm{pp}}/ \mathrm{d}p_\mathrm{T}\mathrm{d}y$ is the invariant cross section in inelastic pp collisions. $\langle T_\mathrm{pPb} \rangle$   = $\langle N_\mathrm{coll} \rangle/ \sigma^{\mathrm{INEL}}$  is the average nuclear overlap function, which accounts for the nuclear collision geometry as obtained from a Glauber model~\cite{Loizides:2017ack}.
 If the nuclear modification factor is unity, then the yield in nuclear collisions is the same as from an incoherent superposition of nucleon--nucleon collisions. \\
  In the absence of  \kstar and  \phim measurements in pp collisions at  $\sqrt{s}$  = 8.16 TeV, the reference\pt spectra are obtained from the distributions measured in pp collisions at  $\sqrt{s}$  = 8 TeV~\cite{Acharya:2019wyb} scaled by the ratio between the \pt spectra  at the two energies as obtained from PYTHIA 8.230~\cite{Sjostrand:2014zea}.  
 For the systematic study the reference pp spectra are also obtained using the measured \pt spectrum at $\sqrt{s}$  = 7 TeV~\cite{Abelev:2012hy}. The total systematic uncertainty of the pp reference spectrum is then calculated as the quadrature sum of the systematic uncertainties of the measured \pt spectrum at $\sqrt{s}$  = 8 TeV and the difference of the reference spectra obtained using the measured \pt spectra at $\sqrt{s}$  = 7 and 8 TeV.   The systematic uncertainties of the reference \pt spectra of \kstar (\phim) are 
 11.5\% (7.3\%) for the  low \pt ($<$ 4 \GeVc) and 15.5\% (7.4\%) for the  high \pt ($>$ 4 \GeVc)~\cite{Adam:2016dau}. 
 The  systematic and statistical uncertainties of  \RpPb are calculated as the quadrature sum of respective uncertainties of the \pt spectra in \pPb and pp collisions. The value of the nucleon--nucleon inelastic cross section for the reference spectra at $\sqrt{s}$  = 8.16 TeV is (72.5$\pm$0.5) mb, taken from Ref.~\cite{Loizides:2017ack}.  \\
  The \RpPb measurements of \kstar, \phim and multi-strange baryon ($\Xi$ and $\Omega$)  in \pPb collisions at \snn  = 5.02 TeV are also reported here. 
  The \RpPb of \kstar and \phim at  \snn  = 5.02 TeV are calculated from the measured \pt spectra in pp and \pPb collisions published in Refs.~\cite{Adam:2016bpr, Acharya:2019qge,ALICE:2021ptz}.
 The \pt spectra measurements of multi-strange baryon ($\Xi$ and $\Omega$) production in \pPb collisions are reported in Ref.~\cite{Adam:2015vsf}.
 \begin{figure}[!h]
   	\begin{center}
             \includegraphics[scale = 0.8]{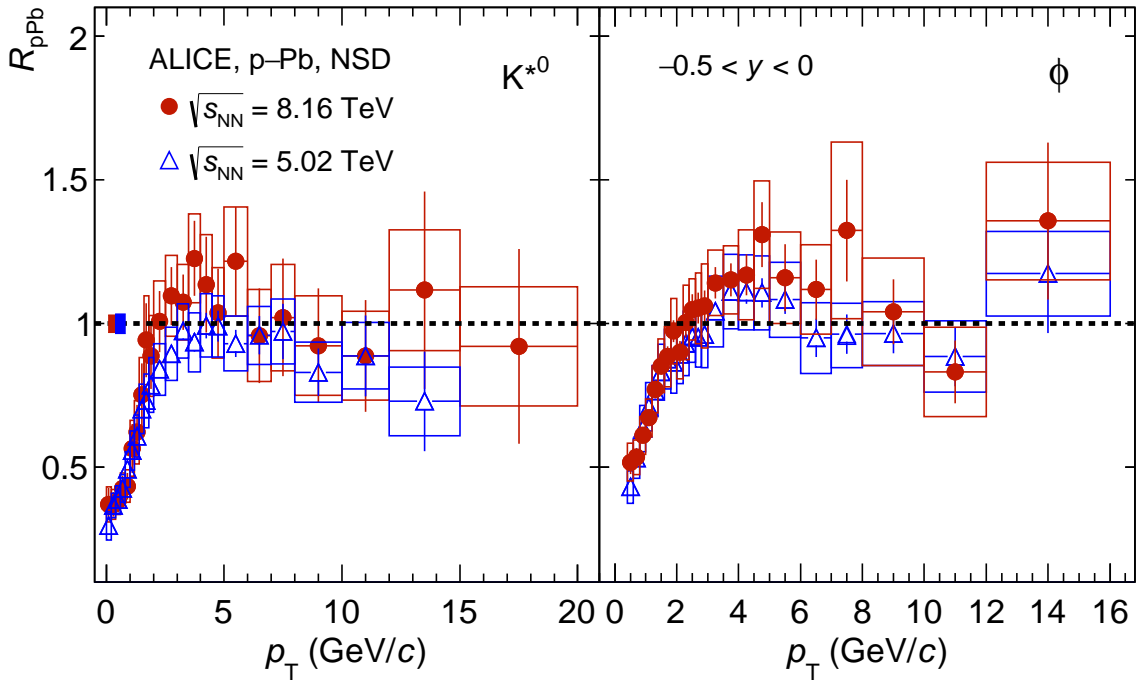}
        \end{center}
   	\caption{Nuclear modification factor of \kstar and \phim as a function of \pt in p--Pb collisions at different energies \snn  = 5.02 and 8.16 TeV. The statistical and systematic  uncertainties are represented by vertical bars and boxes, respectively. The normalization uncertainties are shown in each panel as boxes around \RpPb $=$ 1 near \pt $=$ 0 \GeVc.}
   \label{fig:NMFphiK}
   \end{figure}
Due to the unavailability of multi-strange baryon measurements in pp collisions at $\sqrt{s}$  = 5.02 TeV, reference \pt spectra are calculated by interpolating the measurements at
 $\sqrt{s}$  = 2.76~\cite{ABELEV:2013zaa}  and 7 TeV~\cite{Abelev:2012jp}, in each \pt interval, assuming a power law dependence as a function of $\sqrt{s}$. The systematic uncertainties of the reference \pt spectra are taken as the maximum relative systematic uncertainty of the measured \pt spectra at $\sqrt{s}$  = 2.76 and 7 TeV. This approach is similar to the one as described in Ref.~\cite{Adam:2016dau} to obtain reference \pt spectra for $\pi^{\pm}$, K$^{\pm}$ and p($\overline{\rm p}$) in pp collisions at $\sqrt{s}$  = 5.02 TeV. 
Figure~\ref{fig:NMFphiK} shows the nuclear modification factor of \kstar (left panel) and \phim (right panel) as a function of \pt in \pPb collisions at \snn~ =~5.02 and 8.16 TeV.
 At intermediate \pt (2--8 \GeVc), there is a hint of  increase  in  \RpPb, above unity which is more pronounced for \kstar than for \phim. The measurements are consistent with each other within uncertainties. No significant energy dependence of \RpPb is observed  for resonances in p--Pb collisions at the LHC energies.
   \begin{figure}[H]
	\begin{center}
	\includegraphics[scale = 0.78]{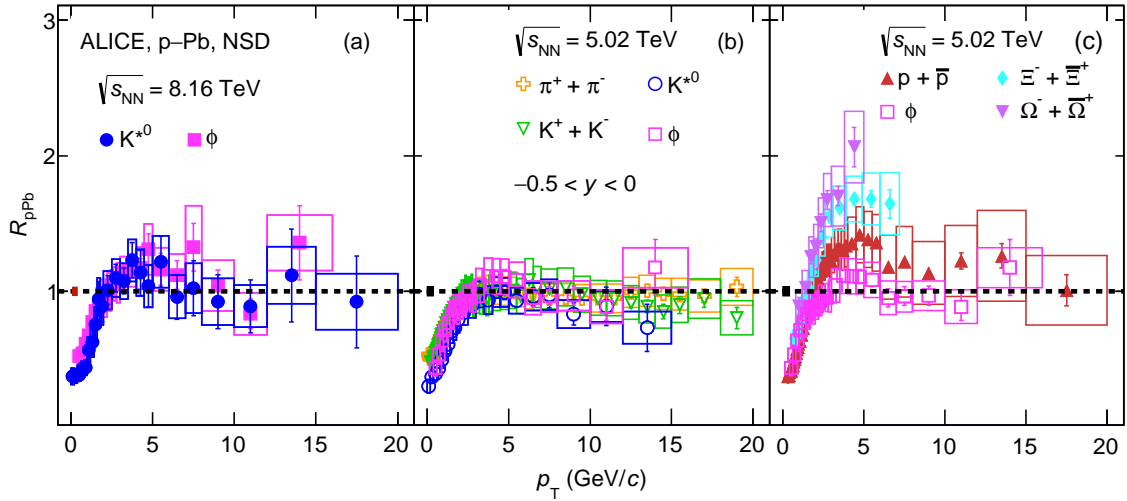}
              \end{center}
              \caption{
            The nuclear modification factor \RpPb as a function of transverse momentum \pt for different particle species in \pPb collisions at \snn  = 5.02  and 8.16 TeV. For comparison the results for $\pi$, K,  and p~\cite{Adam:2016dau} are also shown. 
            The statistical and systematic  uncertainties are represented by vertical bars and boxes, respectively. The normalization uncertainties are shown in each panel as boxes around \RpPb $=$ 1 near \pt $=$ 0 \GeVc.}
\label{fig:NMFAP}
\end{figure} 
		
Figure~\ref{fig:NMFAP} shows the particle species dependence of the nuclear modification factors  in \pPb collisions at \snn  = 5.02 and 8.16 TeV. Panels (a) and (b) show  \RpPb of  \kstar and \phim at \snn  = 8.16 and 5.02 TeV, respectively. Previous measurements of $\pi$ and K  mesons at \mbox{\snn  = 5.02 TeV}~\cite{Adam:2016dau} are also shown in panel (b). Panel (c) shows the  \RpPb of multi-strange baryons ($\Xi$, $\Omega$) at \snn  = 5.02 TeV. To study the mass dependence of baryons and to compare baryons and mesons, the \RpPb of protons taken from~\cite{Adam:2016dau} and that of \phim mesons are also shown in panel (c). At low \pt ($<$ 2 \GeVc), the  \RpPb is less than unity for all hadrons.
   The  measurements of \kstar and \phim  at \snn  = 5.02 and 8.16 TeV are consistent with each other within uncertainties, no flavor dependence in  \RpPb is observed.
   At intermediate \pt (2--8 \GeVc), the  \RpPb  of baryons shows a Cronin-like enhancement above unity~\cite{Kopeliovich:2002yh}. The  \RpPb shows a mass ordering and larger values are observed for the baryons with higher masses. A  similar mass ordering for baryons in this  \pt region is also reported by CMS in Ref.~\cite{Sirunyan:2019rfz} and the results are consistent with a hydrodynamical expectation of the radial flow~\cite{Pierog:2013ria}.
It is also observed that the  \RpPb of \phim meson is smaller than that of the proton in spite of their similar masses, which may indicate baryon-meson ordering. Therefore, along with the presence of a strong radial flow component, there are other effects like different production mechanism for baryons and mesons which affect the  \RpPb in this \pt region. Similar behavior is also observed in \PbPb collisions in this \pt region~\cite{Acharya:2019yoi}. 
At high \pt ($>$ 8 \GeVc), the  \RpPb values of all particles are consistent with unity within the uncertainties in  \pPb collisions at  \snn  = 5.02 and 8.16 TeV which suggests that there is no modification in  \RpPb due to cold-nuclear matter effects for different particle species.  
Similar findings are also reported for $\pi^{0}$  meson with \pt up to 200 \GeVc in p--Pb collisions at \snn = 8.16 TeV~\cite{ALICE:2021est}, for charged hadrons  in p--Pb collisions at \snn  = 5.02 TeV by ALICE~\cite{Adam:2016dau,Acharya:2018qsh}, and for strange hadrons by CMS in p--Pb collisions at  \snn  = 5.02 TeV~\cite{Sirunyan:2019rfz} and by STAR in  d--Au collisions at  \snn  = 200 GeV ~\cite{Abelev:2008yz}.
   
 \section{Summary}
The production of  \kstar and \phim mesons as a function of \pt has been measured in the rapidity interval
$-0.5~< y < $~0 for various multiplicity classes in \pPb collisions at  \snn  = 8.16 TeV with  the ALICE detector.
The EPOS-LHC model describes the NSD \pt distribution while the DPMJET and HIJING models largely overestimate the distribution at low \pt. 
A significant evolution of spectral shapes with multiplicity is observed for \pt $<$ 4 \GeVc, with a pattern similar to that of Pb--Pb collisions, which can be attributed to the collective radial expansion of the system. The spectral shapes are similar for all multiplicity classes at high \pt. 
The scaled \pt-integrated yields $((\dNdyy)/(\langle\mathrm{d}N_{\mathrm{ch}}/\mathrm{d}\eta\rangle_{|\eta|<0.5}))$
as a function of multiplicity show a smooth evolution from small systems, pp and p--Pb, to Pb--Pb, and the values are similar for a given multiplicity, irrespective of the colliding systems and energies, suggesting that the hadrochemistry at LHC energies is mainly driven by the event multiplicity. 
The \meanpt values of \kstar and \phim  increase as a function of multiplicity and  follow a different trend in p--Pb and pp than Pb--Pb collisions. 
The EPOS-LHC model which includes parameterized flow gives a good quantitative description of the scaled \pt-integrated yields and describes qualitatively the increase in \meanpt values with multiplicity for both \kstar and \phim.
An empirical $x_{\rm{T}}$ scaling for \kstar and \phim holds (within roughly 20\%) in the hard scattering region of the particle production. The obtained value of the exponent ($n$ $\approx$ 5) is lower than at RHIC energies which suggests an increasing contribution of hard scattering processes at higher \snn. 
Furthermore, the value of the exponent $n$ in \pPb collisions is compatible with those in pp collisions for $\pi^{\pm}$, K$^{\pm}$ and \kstar suggesting that the  high--\pt particle production mechanism is similar in both  collision systems.
No significant energy dependence in  \RpPb is observed for  \kstar and  \phim in \pPb collisions at \mbox{\snn  = 5.02 and 8.16 TeV}. 
At intermediate \pt (2 $<$ \pt $<$ 8 GeV/$c$),  \RpPb values for multi-strange baryon ($\Xi$ and $\Omega$) and the  protons in \pPb collisions at \snn  = 5.02 TeV show a Cronin-like enhancement and the values are found to be significantly larger  than those for $\pi^{\pm}$, K$^{\pm}$, \kstar and \phim.
The  \RpPb values are consistent with unity within the uncertainties for all species at $\pt >$ 8 \GeVc, which further confirms the absence of cold--nuclear matter effects in this \pt range.  
 Future measurements of light flavor hadron ($\pi^{\pm}$, K$^{\pm}$, p($\bar{\rm{p}}$) etc.) yields up to high \pt in \pPb collisions at \snn $=$ 8.16 TeV are required for a comprehensive study of nuclear modification factor and $x_{\rm{T}}$ scaling.

\newenvironment{acknowledgement}{\relax}{\relax}
\begin{acknowledgement}
\section*{Acknowledgements}

The ALICE Collaboration would like to thank all its engineers and technicians for their invaluable contributions to the construction of the experiment and the CERN accelerator teams for the outstanding performance of the LHC complex.
The ALICE Collaboration gratefully acknowledges the resources and support provided by all Grid centres and the Worldwide LHC Computing Grid (WLCG) collaboration.
The ALICE Collaboration acknowledges the following funding agencies for their support in building and running the ALICE detector:
A. I. Alikhanyan National Science Laboratory (Yerevan Physics Institute) Foundation (ANSL), State Committee of Science and World Federation of Scientists (WFS), Armenia;
Austrian Academy of Sciences, Austrian Science Fund (FWF): [M 2467-N36] and Nationalstiftung f\"{u}r Forschung, Technologie und Entwicklung, Austria;
Ministry of Communications and High Technologies, National Nuclear Research Center, Azerbaijan;
Conselho Nacional de Desenvolvimento Cient\'{\i}fico e Tecnol\'{o}gico (CNPq), Financiadora de Estudos e Projetos (Finep), Funda\c{c}\~{a}o de Amparo \`{a} Pesquisa do Estado de S\~{a}o Paulo (FAPESP) and Universidade Federal do Rio Grande do Sul (UFRGS), Brazil;
Ministry of Education of China (MOEC) , Ministry of Science \& Technology of China (MSTC) and National Natural Science Foundation of China (NSFC), China;
Ministry of Science and Education and Croatian Science Foundation, Croatia;
Centro de Aplicaciones Tecnol\'{o}gicas y Desarrollo Nuclear (CEADEN), Cubaenerg\'{\i}a, Cuba;
Ministry of Education, Youth and Sports of the Czech Republic, Czech Republic;
The Danish Council for Independent Research | Natural Sciences, the VILLUM FONDEN and Danish National Research Foundation (DNRF), Denmark;
Helsinki Institute of Physics (HIP), Finland;
Commissariat \`{a} l'Energie Atomique (CEA) and Institut National de Physique Nucl\'{e}aire et de Physique des Particules (IN2P3) and Centre National de la Recherche Scientifique (CNRS), France;
Bundesministerium f\"{u}r Bildung und Forschung (BMBF) and GSI Helmholtzzentrum f\"{u}r Schwerionenforschung GmbH, Germany;
General Secretariat for Research and Technology, Ministry of Education, Research and Religions, Greece;
National Research, Development and Innovation Office, Hungary;
Department of Atomic Energy Government of India (DAE), Department of Science and Technology, Government of India (DST), University Grants Commission, Government of India (UGC) and Council of Scientific and Industrial Research (CSIR), India;
Indonesian Institute of Science, Indonesia;
Istituto Nazionale di Fisica Nucleare (INFN), Italy;
Japanese Ministry of Education, Culture, Sports, Science and Technology (MEXT), Japan Society for the Promotion of Science (JSPS) KAKENHI and Japanese Ministry of Education, Culture, Sports, Science and Technology (MEXT)of Applied Science (IIST), Japan;
Consejo Nacional de Ciencia (CONACYT) y Tecnolog\'{i}a, through Fondo de Cooperaci\'{o}n Internacional en Ciencia y Tecnolog\'{i}a (FONCICYT) and Direcci\'{o}n General de Asuntos del Personal Academico (DGAPA), Mexico;
Nederlandse Organisatie voor Wetenschappelijk Onderzoek (NWO), Netherlands;
The Research Council of Norway, Norway;
Commission on Science and Technology for Sustainable Development in the South (COMSATS), Pakistan;
Pontificia Universidad Cat\'{o}lica del Per\'{u}, Peru;
Ministry of Education and Science, National Science Centre and WUT ID-UB, Poland;
Korea Institute of Science and Technology Information and National Research Foundation of Korea (NRF), Republic of Korea;
Ministry of Education and Scientific Research, Institute of Atomic Physics and Ministry of Research and Innovation and Institute of Atomic Physics, Romania;
Joint Institute for Nuclear Research (JINR), Ministry of Education and Science of the Russian Federation, National Research Centre Kurchatov Institute, Russian Science Foundation and Russian Foundation for Basic Research, Russia;
Ministry of Education, Science, Research and Sport of the Slovak Republic, Slovakia;
National Research Foundation of South Africa, South Africa;
Swedish Research Council (VR) and Knut \& Alice Wallenberg Foundation (KAW), Sweden;
European Organization for Nuclear Research, Switzerland;
Suranaree University of Technology (SUT), National Science and Technology Development Agency (NSDTA) and Office of the Higher Education Commission under NRU project of Thailand, Thailand;
Turkish Energy, Nuclear and Mineral Research Agency (TENMAK), Turkey;
National Academy of  Sciences of Ukraine, Ukraine;
Science and Technology Facilities Council (STFC), United Kingdom;
National Science Foundation of the United States of America (NSF) and United States Department of Energy, Office of Nuclear Physics (DOE NP), United States of America.
\end{acknowledgement}
\bibliographystyle{utphys}   
\bibliography{KstarPhipPb8p16TeV}

\providecommand{\href}[2]{#2}\begingroup\raggedright\begin{thebibliography}{10}

\bibitem{Braun-Munzinger:2015hba}
P.~Braun-Munzinger, V.~Koch, T.~Sch\"afer, and J.~Stachel, ``{Properties of hot
  and dense matter from relativistic heavy ion collisions}'',
  \href{http://dx.doi.org/10.1016/j.physrep.2015.12.003}{{\em Phys. Rept.}
  {\bfseries 621} (2016) 76--126},
  \href{http://arxiv.org/abs/1510.00442}{{\ttfamily arXiv:1510.00442
  [nucl-th]}}.

\bibitem{Adams:2005dq}
{\bfseries STAR} Collaboration, J.~Adams {\em et~al.}, ``{Experimental and
  theoretical challenges in the search for the quark gluon plasma: The STAR
  Collaboration's critical assessment of the evidence from RHIC collisions}'',
  \href{http://dx.doi.org/10.1016/j.nuclphysa.2005.03.085}{{\em Nucl. Phys.}
  {\bfseries A757} (2005) 102--183},
\href{http://arxiv.org/abs/nucl-ex/0501009}{{\ttfamily arXiv:nucl-ex/0501009
  [nucl-ex]}}.

\bibitem{Schukraft:2011na}
{\bfseries ALICE} Collaboration, J.~Schukraft, ``{Heavy Ion physics with the
  ALICE experiment at the CERN LHC}'',
  \href{http://dx.doi.org/10.1098/rsta.2011.0469}{{\em Phil. Trans. Roy. Soc.
  Lond. A} {\bfseries 370} (2012) 917--932},
  \href{http://arxiv.org/abs/1109.4291}{{\ttfamily arXiv:1109.4291 [hep-ex]}}.

\bibitem{Aggarwal:2010mt}
{\bfseries STAR} Collaboration, M.~M. Aggarwal {\em et~al.}, ``{$K^{*0}$
  production in Cu+Cu and Au+Au collisions at $\sqrt{s_{NN}}$ = 62.4 GeV and
  200 GeV}'', \href{http://dx.doi.org/10.1103/PhysRevC.84.034909}{{\em Phys.
  Rev.} {\bfseries C84} (2011) 034909},
\href{http://arxiv.org/abs/1006.1961}{{\ttfamily arXiv:1006.1961 [nucl-ex]}}.

\bibitem{Adams:2004ep}
{\bfseries STAR} Collaboration, J.~Adams {\em et~al.}, ``{K$^{*}(892)^{0}$
  resonance production in Au+Au and p+p collisions at $\sqrt{s_{_{NN}}}=200$
  GeV at STAR}'', \href{http://dx.doi.org/10.1103/PhysRevC.71.064902}{{\em
  Phys. Rev.} {\bfseries C71} (2005) 064902},
\href{http://arxiv.org/abs/nucl-ex/0412019}{{\ttfamily arXiv:nucl-ex/0412019
  [nucl-ex]}}.

\bibitem{Adler:2002sw}
{\bfseries STAR} Collaboration, C.~Adler {\em et~al.}, ``{K$^{*}(892)^{0}$
  production in relativistic heavy ion collisions at $\sqrt{s_{_{NN}}}=$ 130
  GeV}'', \href{http://dx.doi.org/10.1103/PhysRevC.66.061901}{{\em Phys. Rev.}
  {\bfseries C66} (2002) 061901},
\href{http://arxiv.org/abs/nucl-ex/0205015}{{\ttfamily arXiv:nucl-ex/0205015
  [nucl-ex]}}.

\bibitem{Anticic:2011zr}
{\bfseries NA49} Collaboration, T.~Anticic {\em et~al.}, ``{$K^{\ast}(892)^0$
  and $\bar{K}^{\ast}(892)^0$ production in central Pb+Pb, Si+Si, C+C and
  inelastic p+p collisions at 158$A$~GeV}'',
  \href{http://dx.doi.org/10.1103/PhysRevC.84.064909}{{\em Phys. Rev.}
  {\bfseries C84} (2011) 064909},
\href{http://arxiv.org/abs/1105.3109}{{\ttfamily arXiv:1105.3109 [nucl-ex]}}.

\bibitem{Abelev:2008zk}
{\bfseries STAR} Collaboration, B.~I. Abelev {\em et~al.}, ``{Energy and system
  size dependence of $\phi$ meson production in Cu+Cu and Au+Au collisions}'',
  \href{http://dx.doi.org/10.1016/j.physletb.2009.02.037}{{\em Phys. Lett.}
  {\bfseries B673} (2009) 183--191},
\href{http://arxiv.org/abs/0810.4979}{{\ttfamily arXiv:0810.4979 [nucl-ex]}}.

\bibitem{Abelev:2008aa}
{\bfseries STAR} Collaboration, B.~I. Abelev {\em et~al.}, ``{Measurements of
  $\phi$ meson production in relativistic heavy-ion collisions at RHIC}'',
  \href{http://dx.doi.org/10.1103/PhysRevC.79.064903}{{\em Phys. Rev.}
  {\bfseries C79} (2009) 064903},
\href{http://arxiv.org/abs/0809.4737}{{\ttfamily arXiv:0809.4737 [nucl-ex]}}.

\bibitem{Adamczyk:2015lvo}
{\bfseries STAR} Collaboration, L.~Adamczyk {\em et~al.}, ``{Probing parton
  dynamics of QCD matter with $\Omega$ and $\phi$ production}'',
  \href{http://dx.doi.org/10.1103/PhysRevC.93.021903}{{\em Phys. Rev.}
  {\bfseries C93} (2016) 021903},
\href{http://arxiv.org/abs/1506.07605}{{\ttfamily arXiv:1506.07605 [nucl-ex]}}.

\bibitem{Alt:2008iv}
{\bfseries NA49} Collaboration, C.~Alt {\em et~al.}, ``{Energy dependence of
  $\phi$ meson production in central Pb+Pb collisions at $\sqrt{s_{_{NN}}}$ = 6
  to 17 GeV}'', \href{http://dx.doi.org/10.1103/PhysRevC.78.044907}{{\em Phys.
  Rev.} {\bfseries C78} (2008) 044907},
\href{http://arxiv.org/abs/0806.1937}{{\ttfamily arXiv:0806.1937 [nucl-ex]}}.

\bibitem{Adare:2010pt}
{\bfseries PHENIX} Collaboration, A.~Adare {\em et~al.}, ``{Nuclear
  modification factors of $\phi$ mesons in $d+$Au, Cu+Cu and Au+Au collisions
  at $\sqrt{s_{_{NN}}}=200$ GeV}'',
  \href{http://dx.doi.org/10.1103/PhysRevC.83.024909}{{\em Phys. Rev.}
  {\bfseries C83} (2011) 024909},
\href{http://arxiv.org/abs/1004.3532}{{\ttfamily arXiv:1004.3532 [nucl-ex]}}.

\bibitem{Abelev:2014uua}
{\bfseries ALICE} Collaboration, B.~Abelev {\em et~al.}, ``{$K^*(892)^0$ and
  $\phi(1020)$ production in Pb-Pb collisions at $\sqrt{s_{NN}}$ = 2.76 TeV}'',
  \href{http://dx.doi.org/10.1103/PhysRevC.91.024609}{{\em Phys. Rev.}
  {\bfseries C91} (2015) 024609},
\href{http://arxiv.org/abs/1404.0495}{{\ttfamily arXiv:1404.0495 [nucl-ex]}}.

\bibitem{Adare:2014eyu}
{\bfseries PHENIX} Collaboration, A.~Adare {\em et~al.}, ``{Measurement of
  K$^{*0}$ in $p+p$, $d$+Au, and Cu+Cu collisions at $\sqrt{s_{_{NN}}}$ = 200
  GeV}'', \href{http://dx.doi.org/10.1103/PhysRevC.90.054905}{{\em Phys. Rev.}
  {\bfseries C90} (2014) 054905},
\href{http://arxiv.org/abs/1405.3628}{{\ttfamily arXiv:1405.3628 [nucl-ex]}}.

\bibitem{Adam:2017zbf}
{\bfseries ALICE} Collaboration, J.~Adam {\em et~al.}, ``{K$^{*}(892)^{0}$ and
  $\phi(1020)$ meson production at high transverse momentum in pp and Pb$-$Pb
  collisions at $\sqrt{s_\mathrm{NN}}$ = 2.76 TeV}'',
  \href{http://dx.doi.org/10.1103/PhysRevC.95.064606}{{\em Phys. Rev.}
  {\bfseries C95} (2017) 064606},
\href{http://arxiv.org/abs/1702.00555}{{\ttfamily arXiv:1702.00555 [nucl-ex]}}.

\bibitem{Abelev:2012hy}
{\bfseries ALICE} Collaboration, B.~Abelev {\em et~al.}, ``{Production of
  $K^*(892)^0$ and $\phi(1020)$ in $pp$ collisions at $\sqrt{s}=7$ TeV}'',
  \href{http://dx.doi.org/10.1140/epjc/s10052-012-2183-y}{{\em Eur. Phys. J.}
  {\bfseries C72} (2012) 2183},
\href{http://arxiv.org/abs/1208.5717}{{\ttfamily arXiv:1208.5717 [hep-ex]}}.

\bibitem{ALICE:2017jyt}
{\bfseries ALICE} Collaboration, J.~Adam {\em et~al.}, ``{Enhanced production
  of multi-strange hadrons in high-multiplicity proton-proton collisions}'',
  \href{http://dx.doi.org/10.1038/nphys4111}{{\em Nature Phys.} {\bfseries 13}
  (2017) 535--539},
\href{http://arxiv.org/abs/1606.07424}{{\ttfamily arXiv:1606.07424 [nucl-ex]}}.

\bibitem{Acharya:2019qge}
{\bfseries ALICE} Collaboration, S.~Acharya {\em et~al.}, ``{Evidence of
  rescattering effect in Pb--Pb collisions at the LHC through production of
  $\rm{K}^{*}(892)^{0}$ and $\phi(1020)$ mesons}'',
  \href{http://dx.doi.org/10.1016/j.physletb.2020.135225}{{\em Phys. Lett.}
  {\bfseries B802} (2020) 135225},
\href{http://arxiv.org/abs/1910.14419}{{\ttfamily arXiv:1910.14419 [nucl-ex]}}.

\bibitem{ALICE:2021ptz}
{\bfseries ALICE} Collaboration, S.~Acharya {\em et~al.}, ``{Production of
  K$^{*}(892)^{0}$ and $\phi(1020)$ in pp and Pb--Pb collisions at
  $\sqrt{s_{\rm NN}} = 5.02$ TeV}'',
  \href{http://dx.doi.org/10.1103/PhysRevC.106.034907}{{\em Phys. Rev. C}
  {\bfseries 106} (2022) 034907},
  \href{http://arxiv.org/abs/2106.13113}{{\ttfamily arXiv:2106.13113
  [nucl-ex]}}.

\bibitem{Acharya:2018orn}
{\bfseries ALICE} Collaboration, S.~Acharya {\em et~al.}, ``{Multiplicity
  dependence of light-flavor hadron production in pp collisions at $\sqrt{s}$ =
  7 TeV}'', \href{http://dx.doi.org/10.1103/PhysRevC.99.024906}{{\em Phys.
  Rev.} {\bfseries C99} (2019) 024906},
\href{http://arxiv.org/abs/1807.11321}{{\ttfamily arXiv:1807.11321 [nucl-ex]}}.

\bibitem{Acharya:2019wyb}
{\bfseries ALICE} Collaboration, S.~Acharya {\em et~al.},
  ``{$\rm{K}^{*}(\rm{892})^{0}$ and $\phi(1020)$ production at midrapidity in
  pp collisions at $\sqrt{s}$ = 8 TeV}'',
  \href{http://dx.doi.org/10.1103/PhysRevC.102.024912}{{\em Phys. Rev.}
  {\bfseries C102} (2020) 024912},
\href{http://arxiv.org/abs/1910.14410}{{\ttfamily arXiv:1910.14410 [nucl-ex]}}.

\bibitem{Acharya:2019bli}
{\bfseries ALICE} Collaboration, S.~Acharya {\em et~al.}, ``{Multiplicity
  dependence of K*(892)$^{0}$ and $\phi$(1020) production in pp collisions at
  $\sqrt {s}$ = 13 TeV}'',
  \href{http://dx.doi.org/10.1016/j.physletb.2020.135501}{{\em Phys. Lett.}
  {\bfseries B807} (2020) 135501},
\href{http://arxiv.org/abs/1910.14397}{{\ttfamily arXiv:1910.14397 [nucl-ex]}}.

\bibitem{Adam:2016bpr}
{\bfseries ALICE} Collaboration, J.~Adam {\em et~al.}, ``{Production of K$^{*}$
  (892)$^{0}$ and $\phi $ (1020) in p--Pb collisions at $\sqrt{s_{{\text
  {NN}}}}$ = 5.02 TeV}'',
  \href{http://dx.doi.org/10.1140/epjc/s10052-016-4088-7}{{\em Eur. Phys. J.}
  {\bfseries C76} (2016) 245},
\href{http://arxiv.org/abs/1601.07868}{{\ttfamily arXiv:1601.07868 [nucl-ex]}}.

\bibitem{Abelev:2012sk}
{\bfseries ALICE} Collaboration, B.~Abelev {\em et~al.}, ``{Transverse
  sphericity of primary charged particles in minimum bias proton-proton
  collisions at $\sqrt{s}=0.9$, 2.76 and 7 TeV}'',
  \href{http://dx.doi.org/10.1140/epjc/s10052-012-2124-9}{{\em Eur. Phys. J.}
  {\bfseries C72} (2012) 2124},
\href{http://arxiv.org/abs/1205.3963}{{\ttfamily arXiv:1205.3963 [hep-ex]}}.

\bibitem{Abelev:2008yz}
{\bfseries STAR} Collaboration, B.~I. Abelev {\em et~al.}, ``{Hadronic
  resonance production in d+Au collisions at $\sqrt{s_\mathrm{NN}}$ = 200 GeV
  at RHIC}'', \href{http://dx.doi.org/10.1103/PhysRevC.78.044906}{{\em Phys.
  Rev.} {\bfseries C78} (2008) 044906},
\href{http://arxiv.org/abs/0801.0450}{{\ttfamily arXiv:0801.0450 [nucl-ex]}}.

\bibitem{Adam:2016dau}
{\bfseries ALICE} Collaboration, J.~Adam {\em et~al.}, ``{Multiplicity
  dependence of charged pion, kaon, and (anti)proton production at large
  transverse momentum in p-Pb collisions at $\mathbf{\sqrt{{\textit s}_{\rm
  NN}}}$ = 5.02 TeV}'',
  \href{http://dx.doi.org/10.1016/j.physletb.2016.07.050}{{\em Phys. Lett.}
  {\bfseries B760} (2016) 720--735},
\href{http://arxiv.org/abs/1601.03658}{{\ttfamily arXiv:1601.03658 [nucl-ex]}}.

\bibitem{Adam:2015vsf}
{\bfseries ALICE} Collaboration, J.~Adam {\em et~al.}, ``{Multi-strange baryon
  production in p-Pb collisions at $\sqrt{s_\mathbf{NN}}=5.02$ TeV}'',
  \href{http://dx.doi.org/10.1016/j.physletb.2016.05.027}{{\em Phys. Lett.}
  {\bfseries B758} (2016) 389--401},
\href{http://arxiv.org/abs/1512.07227}{{\ttfamily arXiv:1512.07227 [nucl-ex]}}.

\bibitem{ALICE:2019smg}
{\bfseries ALICE} Collaboration, S.~Acharya {\em et~al.}, ``{Measurement of
  $\Lambda$(1520) production in pp collisions at $\sqrt{s}$ = 7 TeV and p-Pb
  collisions at $\sqrt{s_{\rm{NN}}}$ = 5.02 TeV}'',
  \href{http://dx.doi.org/10.1140/epjc/s10052-020-7687-2}{{\em Eur. Phys. J.}
  {\bfseries C80} (2020) 160},
\href{http://arxiv.org/abs/1909.00486}{{\ttfamily arXiv:1909.00486 [nucl-ex]}}.

\bibitem{ALICE:2017pgw}
{\bfseries ALICE} Collaboration, D.~Adamova {\em et~al.}, ``{Production of
  $\Sigma(1385)^{\pm}$ and $\Xi(1530)^{0}$ in p-Pb collisions at $\sqrt{s_{\rm
  NN}}=5.02$ TeV}'',
  \href{http://dx.doi.org/10.1140/epjc/s10052-017-4943-1}{{\em Eur. Phys. J. C}
  {\bfseries 77} (2017) 389}, \href{http://arxiv.org/abs/1701.07797}{{\ttfamily
  arXiv:1701.07797 [nucl-ex]}}.

\bibitem{ALICE:2014zxz}
{\bfseries ALICE} Collaboration, B.~B. Abelev {\em et~al.}, ``{Production of
  $\Sigma(1385)^{\pm}$ and $\Xi(1530)^{0}$ in proton-proton collisions at
  $\sqrt{s}=$ 7 TeV}'',
  \href{http://dx.doi.org/10.1140/epjc/s10052-014-3191-x}{{\em Eur. Phys. J. C}
  {\bfseries 75} (2015) 1}, \href{http://arxiv.org/abs/1406.3206}{{\ttfamily
  arXiv:1406.3206 [nucl-ex]}}.

\bibitem{Pierog:2013ria}
T.~Pierog, I.~Karpenko, J.~M. Katzy, E.~Yatsenko, and K.~Werner, ``{EPOS LHC:
  Test of collective hadronization with data measured at the CERN Large Hadron
  Collider}'', \href{http://dx.doi.org/10.1103/PhysRevC.92.034906}{{\em Phys.
  Rev.} {\bfseries C92} (2015) 034906},
\href{http://arxiv.org/abs/1306.0121}{{\ttfamily arXiv:1306.0121 [hep-ph]}}.

\bibitem{Roesler:2000he}
S.~Roesler, R.~Engel, and J.~Ranft,
  \href{http://dx.doi.org/10.1007/978-3-642-18211-2_166}{``{The Monte Carlo
  event generator DPMJET-III}'',} in {\em {International Conference on Advanced
  Monte Carlo for Radiation Physics, Particle Transport Simulation and
  Applications (MC 2000)}}, pp.~1033--1038.
\newblock 12, 2000.
\newblock \href{http://arxiv.org/abs/hep-ph/0012252}{{\ttfamily
  arXiv:hep-ph/0012252}}.

\bibitem{Gyulassy:1994ew}
M.~Gyulassy and X.-N. Wang, ``{HIJING 1.0: A Monte Carlo program for parton and
  particle production in high-energy hadronic and nuclear collisions}'',
  \href{http://dx.doi.org/10.1016/0010-4655(94)90057-4}{{\em Comput. Phys.
  Commun.} {\bfseries 83} (1994) 307},
\href{http://arxiv.org/abs/nucl-th/9502021}{{\ttfamily arXiv:nucl-th/9502021
  [nucl-th]}}.

\bibitem{Acharya:2019kyh}
{\bfseries ALICE} Collaboration, S.~Acharya {\em et~al.}, ``{Multiplicity
  dependence of (multi-)strange hadron production in proton-proton collisions
  at $\sqrt{s}$ = 13 TeV}'',
  \href{http://dx.doi.org/10.1140/epjc/s10052-020-7673-8}{{\em Eur. Phys. J.}
  {\bfseries C80} (2020) 167},
\href{http://arxiv.org/abs/1908.01861}{{\ttfamily arXiv:1908.01861 [nucl-ex]}}.

\bibitem{Abelev:2013vea}
{\bfseries ALICE} Collaboration, B.~Abelev {\em et~al.}, ``{Centrality
  dependence of $\pi$, K, p production in Pb-Pb collisions at $\sqrt{s_{NN}}$ =
  2.76 TeV}'', \href{http://dx.doi.org/10.1103/PhysRevC.88.044910}{{\em Phys.
  Rev.} {\bfseries C88} (2013) 044910},
\href{http://arxiv.org/abs/1303.0737}{{\ttfamily arXiv:1303.0737 [hep-ex]}}.

\bibitem{Acharya:2019yoi}
{\bfseries ALICE} Collaboration, S.~Acharya {\em et~al.}, ``{Production of
  charged pions, kaons, and (anti-)protons in Pb-Pb and inelastic $pp$
  collisions at $\sqrt {s_{NN}}$ = 5.02 TeV}'',
  \href{http://dx.doi.org/10.1103/PhysRevC.101.044907}{{\em Phys. Rev.}
  {\bfseries C101} (2020) 044907},
\href{http://arxiv.org/abs/1910.07678}{{\ttfamily arXiv:1910.07678 [nucl-ex]}}.

\bibitem{Collins:1989gx}
J.~C. Collins, D.~E. Soper, and G.~F. Sterman, ``{Factorization of Hard
  Processes in QCD}'', \href{http://dx.doi.org/10.1142/9789814503266_0001}{{\em
  Adv. Ser. Direct. High Energy Phys.} {\bfseries 5} (1989) 1--91},
  \href{http://arxiv.org/abs/hep-ph/0409313}{{\ttfamily arXiv:hep-ph/0409313}}.

\bibitem{deFlorian:2014xna}
D.~de~Florian, R.~Sassot, M.~Epele, R.~J. Hernández-Pinto, and M.~Stratmann,
  ``{Parton-to-Pion Fragmentation Reloaded}'',
  \href{http://dx.doi.org/10.1103/PhysRevD.91.014035}{{\em Phys. Rev.}
  {\bfseries D91} (2015) 014035},
\href{http://arxiv.org/abs/1410.6027}{{\ttfamily arXiv:1410.6027 [hep-ph]}}.

\bibitem{Brodsky:2005fza}
S.~J. Brodsky, H.~J. Pirner, and J.~Raufeisen, ``{Scaling properties of high
  $p_{\rm{T}}$ inclusive hadron production}'',
  \href{http://dx.doi.org/10.1016/j.physletb.2006.03.079}{{\em Phys. Lett. B}
  {\bfseries 637} (2006) 58--63},
  \href{http://arxiv.org/abs/hep-ph/0510315}{{\ttfamily arXiv:hep-ph/0510315}}.

\bibitem{Arleo:2009ch}
F.~Arleo, S.~J. Brodsky, D.~S. Hwang, and A.~M. Sickles, ``{Higher-Twist
  Dynamics in Large Transverse Momentum Hadron Production}'',
  \href{http://dx.doi.org/10.1103/PhysRevLett.105.062002}{{\em Phys. Rev.
  Lett.} {\bfseries 105} (2010) 062002},
\href{http://arxiv.org/abs/0911.4604}{{\ttfamily arXiv:0911.4604 [hep-ph]}}.

\bibitem{Aaltonen:2009ne}
{\bfseries CDF} Collaboration, T.~Aaltonen {\em et~al.}, ``{Measurement of
  Particle Production and Inclusive Differential Cross Sections in $p \bar{p}$
  Collisions at $\sqrt{s}$ = 1.96 TeV}'',
  \href{http://dx.doi.org/10.1103/PhysRevD.82.119903,
  10.1103/PhysRevD.79.112005}{{\em Phys. Rev.} {\bfseries D79} (2009) 112005},
  \href{http://arxiv.org/abs/0904.1098}{{\ttfamily arXiv:0904.1098 [hep-ex]}}.
[Erratum: Phys. Rev.D82,119903(2010)].

\bibitem{CDF:2001hmt}
{\bfseries CDF} Collaboration, D.~Acosta {\em et~al.}, ``{Soft and Hard
  Interactions in $p\bar{p}$ Collisions at $\sqrt{s}=$ 1800 GeV and 630 GeV}'',
  \href{http://dx.doi.org/10.1103/PhysRevD.65.072005}{{\em Phys. Rev. D}
  {\bfseries 65} (2002) 072005}.

\bibitem{Abe:1988yu}
{\bfseries CDF} Collaboration, F.~Abe {\em et~al.}, ``{Transverse Momentum
  Distributions of Charged Particles Produced in $\bar{p}p$ Interactions at
  $\sqrt{s}$ = 630 and 1800 GeV}'',
\href{http://dx.doi.org/10.1103/PhysRevLett.61.1819}{{\em Phys. Rev. Lett.}
  {\bfseries 61} (1988) 1819}.

\bibitem{Albajar:1989an}
{\bfseries UA1} Collaboration, C.~Albajar {\em et~al.}, ``{A Study of the
  General Characteristics of $p\bar{p}$ Collisions at $\sqrt{s}$ = 0.2 TeV to
  0.9 TeV}'',
\href{http://dx.doi.org/10.1016/0550-3213(90)90493-W}{{\em Nucl. Phys.}
  {\bfseries B335} (1990) 261--287}.

\bibitem{Adams:2006nd}
{\bfseries STAR} Collaboration, J.~Adams {\em et~al.}, ``{Identified hadron
  spectra at large transverse momentum in p+p and d+Au collisions at
  $\sqrt{s_{NN}}$ = 200 GeV}'',
  \href{http://dx.doi.org/10.1016/j.physletb.2006.04.032}{{\em Phys. Lett.}
  {\bfseries B637} (2006) 161--169},
\href{http://arxiv.org/abs/nucl-ex/0601033}{{\ttfamily arXiv:nucl-ex/0601033
  [nucl-ex]}}.

\bibitem{ALICE:2020jsh}
{\bfseries ALICE} Collaboration, S.~Acharya {\em et~al.}, ``{Production of
  light-flavor hadrons in pp collisions at $\sqrt{s}$ = 7 and $\sqrt{s}$ = 13
  TeV}'', \href{http://dx.doi.org/10.1140/epjc/s10052-020-08690-5}{{\em Eur.
  Phys. J. C} {\bfseries 81} (2021) 256},
  \href{http://arxiv.org/abs/2005.11120}{{\ttfamily arXiv:2005.11120
  [nucl-ex]}}.

\bibitem{Chatrchyan:2011av}
{\bfseries CMS} Collaboration, S.~Chatrchyan {\em et~al.}, ``{Charged particle
  transverse momentum spectra in $pp$ collisions at $\sqrt{s} = 0.9$ and 7
  TeV}'', \href{http://dx.doi.org/10.1007/JHEP08(2011)086}{{\em JHEP}
  {\bfseries 08} (2011) 086},
\href{http://arxiv.org/abs/1104.3547}{{\ttfamily arXiv:1104.3547 [hep-ex]}}.

\bibitem{Acharya:2018qsh}
{\bfseries ALICE} Collaboration, S.~Acharya {\em et~al.}, ``{Transverse
  momentum spectra and nuclear modification factors of charged particles in pp,
  p-Pb and Pb-Pb collisions at the LHC}'',
  \href{http://dx.doi.org/10.1007/JHEP11(2018)013}{{\em JHEP} {\bfseries 11}
  (2018) 013},
\href{http://arxiv.org/abs/1802.09145}{{\ttfamily arXiv:1802.09145 [nucl-ex]}}.

\bibitem{Sirunyan:2019rfz}
{\bfseries CMS} Collaboration, A.~M. Sirunyan {\em et~al.}, ``{Strange hadron
  production in pp and pPb collisions at $\sqrt{s_\mathrm{NN}}= $ 5.02 TeV}'',
  \href{http://dx.doi.org/10.1103/PhysRevC.101.064906}{{\em Phys. Rev. C}
  {\bfseries 101} (2020) 064906},
  \href{http://arxiv.org/abs/1910.04812}{{\ttfamily arXiv:1910.04812
  [hep-ex]}}.

\bibitem{CMS:2016zzh}
{\bfseries CMS} Collaboration, V.~Khachatryan {\em et~al.}, ``{Multiplicity and
  rapidity dependence of strange hadron production in pp, pPb, and PbPb
  collisions at the LHC}'',
  \href{http://dx.doi.org/10.1016/j.physletb.2017.01.075}{{\em Phys. Lett. B}
  {\bfseries 768} (2017) 103--129},
  \href{http://arxiv.org/abs/1605.06699}{{\ttfamily arXiv:1605.06699
  [nucl-ex]}}.

\bibitem{ALICE:2013wgn}
{\bfseries ALICE} Collaboration, B.~B. Abelev {\em et~al.}, ``{Multiplicity
  Dependence of Pion, Kaon, Proton and Lambda Production in p-Pb Collisions at
  $\sqrt{s_{NN}}$ = 5.02 TeV}'',
  \href{http://dx.doi.org/10.1016/j.physletb.2013.11.020}{{\em Phys. Lett. B}
  {\bfseries 728} (2014) 25--38},
  \href{http://arxiv.org/abs/1307.6796}{{\ttfamily arXiv:1307.6796 [nucl-ex]}}.

\bibitem{Sjostrand:2014zea}
T.~Sj\"ostrand, S.~Ask, J.~R. Christiansen, R.~Corke, N.~Desai, P.~Ilten,
  S.~Mrenna, S.~Prestel, C.~O. Rasmussen, and P.~Z. Skands, ``{An introduction
  to PYTHIA 8.2}'', \href{http://dx.doi.org/10.1016/j.cpc.2015.01.024}{{\em
  Comput. Phys. Commun.} {\bfseries 191} (2015) 159--177},
  \href{http://arxiv.org/abs/1410.3012}{{\ttfamily arXiv:1410.3012 [hep-ph]}}.

\bibitem{ALICE:2021est}
{\bfseries ALICE} Collaboration, S.~Acharya {\em et~al.}, ``{Nuclear
  modification factor of light neutral-meson spectra up to high transverse
  momentum in p--Pb collisions at $\sqrt{s_{\rm NN}}$ = 8.16 TeV}'',
  \href{http://dx.doi.org/10.1016/j.physletb.2022.136943}{{\em Phys. Lett. B}
  {\bfseries 827} (2022) 136943},
  \href{http://arxiv.org/abs/2104.03116}{{\ttfamily arXiv:2104.03116
  [nucl-ex]}}.

\bibitem{Fries:2003vb}
R.~J. Fries, B.~Muller, C.~Nonaka, and S.~A. Bass, ``{Hadronization in heavy
  ion collisions: Recombination and fragmentation of partons}'',
  \href{http://dx.doi.org/10.1103/PhysRevLett.90.202303}{{\em Phys. Rev. Lett.}
  {\bfseries 90} (2003) 202303},
\href{http://arxiv.org/abs/nucl-th/0301087}{{\ttfamily arXiv:nucl-th/0301087
  [nucl-th]}}.

\bibitem{Kang:2012kc}
Z.-B. Kang, I.~Vitev, and H.~Xing, ``{Nuclear modification of high transverse
  momentum particle production in p+A collisions at RHIC and LHC}'',
  \href{http://dx.doi.org/10.1016/j.physletb.2012.10.046}{{\em Phys. Lett.}
  {\bfseries B718} (2012) 482--487},
\href{http://arxiv.org/abs/1209.6030}{{\ttfamily arXiv:1209.6030 [hep-ph]}}.

\bibitem{Kopeliovich:2002yh}
B.~Z. Kopeliovich, J.~Nemchik, A.~Schafer, and A.~V. Tarasov, ``{Cronin effect
  in hadron production off nuclei}'',
  \href{http://dx.doi.org/10.1103/PhysRevLett.88.232303}{{\em Phys. Rev. Lett.}
  {\bfseries 88} (2002) 232303},
  \href{http://arxiv.org/abs/hep-ph/0201010}{{\ttfamily arXiv:hep-ph/0201010}}.

\bibitem{Cronin:1973fd}
J.~W. Cronin, H.~J. Frisch, M.~J. Shochet, J.~P. Boymond, P.~A. Piroue, and
  R.~L. Sumner, ``{Production of Hadrons with Large Transverse Momentum at 200
  GeV and 300 GeV.}'',
\href{http://dx.doi.org/10.1103/PhysRevLett.31.1426}{{\em Phys. Rev. Lett.}
  {\bfseries 31} (1973) 1426--1429}.

\bibitem{Cronin:1974zm}
J.~W. Cronin, H.~J. Frisch, M.~J. Shochet, J.~P. Boymond, R.~Mermod, P.~A.
  Piroue, and R.~L. Sumner, ``{Production of hadrons with large transverse
  momentum at 200, 300, and 400 GeV}'',
  \href{http://dx.doi.org/10.1103/PhysRevD.11.3105}{{\em Phys. Rev. D}
  {\bfseries 11} (1975) 3105--3123}.

\bibitem{Aamodt:2008zz}
{\bfseries ALICE} Collaboration, K.~Aamodt {\em et~al.}, ``{The ALICE
  experiment at the CERN LHC}'',
\href{http://dx.doi.org/10.1088/1748-0221/3/08/S08002}{{\em JINST} {\bfseries
  3} (2008) S08002}.

\bibitem{Abelev:2014ffa}
{\bfseries ALICE} Collaboration, B.~Abelev {\em et~al.}, ``{Performance of the
  ALICE Experiment at the CERN LHC}'',
  \href{http://dx.doi.org/10.1142/S0217751X14300440}{{\em Int. J. Mod. Phys.}
  {\bfseries A29} (2014) 1430044},
\href{http://arxiv.org/abs/1402.4476}{{\ttfamily arXiv:1402.4476 [nucl-ex]}}.

\bibitem{Abbas:2013taa}
{\bfseries ALICE} Collaboration, E.~Abbas {\em et~al.}, ``{Performance of the
  ALICE VZERO system}'',
  \href{http://dx.doi.org/10.1088/1748-0221/8/10/P10016}{{\em JINST} {\bfseries
  8} (2013) P10016},
\href{http://arxiv.org/abs/1306.3130}{{\ttfamily arXiv:1306.3130 [nucl-ex]}}.

\bibitem{Acharya:2018egz}
{\bfseries ALICE} Collaboration, S.~Acharya {\em et~al.}, ``{Charged-particle
  pseudorapidity density at mid-rapidity in p-Pb collisions at
  $\sqrt{s_{\rm{NN}}}$ = 8.16 TeV}'',
  \href{http://dx.doi.org/10.1140/epjc/s10052-019-6801-9}{{\em Eur. Phys. J.}
  {\bfseries C79} (2019) 307},
\href{http://arxiv.org/abs/1812.01312}{{\ttfamily arXiv:1812.01312 [nucl-ex]}}.

\bibitem{Aamodt:2010aa}
{\bfseries ALICE} Collaboration, K.~Aamodt {\em et~al.}, ``{Alignment of the
  ALICE Inner Tracking System with cosmic-ray tracks}'',
  \href{http://dx.doi.org/10.1088/1748-0221/5/03/P03003}{{\em JINST} {\bfseries
  5} (2010) P03003},
\href{http://arxiv.org/abs/1001.0502}{{\ttfamily arXiv:1001.0502
  [physics.ins-det]}}.

\bibitem{Alme:2010ke}
J.~Alme {\em et~al.}, ``{The ALICE TPC, a large 3-dimensional tracking device
  with fast readout for ultra-high multiplicity events}'',
  \href{http://dx.doi.org/10.1016/j.nima.2010.04.042}{{\em Nucl. Instrum.
  Meth.} {\bfseries A622} (2010) 316--367},
\href{http://arxiv.org/abs/1001.1950}{{\ttfamily arXiv:1001.1950
  [physics.ins-det]}}.

\bibitem{ALICE:2015ial}
{\bfseries ALICE} Collaboration, J.~Adam {\em et~al.}, ``{Measurement of pion,
  kaon and proton production in proton\textendash{}proton collisions at
  $\sqrt{s} = 7$ TeV}'',
  \href{http://dx.doi.org/10.1140/epjc/s10052-015-3422-9}{{\em Eur. Phys. J. C}
  {\bfseries 75} (2015) 226}, \href{http://arxiv.org/abs/1504.00024}{{\ttfamily
  arXiv:1504.00024 [nucl-ex]}}.

\bibitem{Akindinov:2013tea}
A.~Akindinov {\em et~al.}, ``{Performance of the ALICE Time-Of-Flight detector
  at the LHC}'', \href{http://dx.doi.org/10.1140/epjp/i2013-13044-x}{{\em Eur.
  Phys. J. Plus} {\bfseries 128} (2013) 44}.

\bibitem{Tanabashi:2018oca}
{\bfseries Particle Data Group} Collaboration, M.~Tanabashi {\em et~al.},
  ``{Review of Particle Physics}'',
\href{http://dx.doi.org/10.1103/PhysRevD.98.030001}{{\em Phys. Rev.} {\bfseries
  D98} (2018) 030001}.

\bibitem{Brun:1119728}
R.~Brun, F.~Bruyant, M.~Maire, A.~C. McPherson, and P.~Zanarini, {\em {GEANT 3:
  user's guide Geant 3.10, Geant 3.11; rev. version}}.
\newblock CERN, Geneva, 1987.
\newblock \url{https://cds.cern.ch/record/1119728}.

\bibitem{Werner:2013tya}
K.~Werner, B.~Guiot, I.~Karpenko, and T.~Pierog, ``{Analyzing radial flow
  features in $p$-Pb and $p-p$ collisions at several TeV by studying
  identified-particle production with the event generator EPOS3}'',
  \href{http://dx.doi.org/10.1103/PhysRevC.89.064903}{{\em Phys. Rev.}
  {\bfseries C89} (2014) 064903},
\href{http://arxiv.org/abs/1312.1233}{{\ttfamily arXiv:1312.1233 [nucl-th]}}.

\bibitem{Werner:2013yia}
K.~Werner, I.~Karpenko, M.~Bleicher, and T.~Pierog, ``{The Physics of EPOS}'',
\href{http://dx.doi.org/10.1051/epjconf/20125205001}{{\em EPJ Web Conf.}
  {\bfseries 52} (2013) 05001}.

\bibitem{Drescher:2000ha}
H.~J. Drescher, M.~Hladik, S.~Ostapchenko, T.~Pierog, and K.~Werner,
  ``{Parton-based Gribov-Regge theory}'',
  \href{http://dx.doi.org/10.1016/S0370-1573(00)00122-8}{{\em Phys. Rept.}
  {\bfseries 350} (2001) 93--289},
\href{http://arxiv.org/abs/hep-ph/0007198}{{\ttfamily arXiv:hep-ph/0007198
  [hep-ph]}}.

\bibitem{Werner:2007bf}
K.~Werner, ``{Core-corona separation in ultrarelativistic heavy ion
  collisions}'', \href{http://dx.doi.org/10.1103/PhysRevLett.98.152301}{{\em
  Phys. Rev. Lett.} {\bfseries 98} (2007) 152301},
\href{http://arxiv.org/abs/0704.1270}{{\ttfamily arXiv:0704.1270 [nucl-th]}}.

\bibitem{Pierog:2009zt}
T.~Pierog and K.~Werner, ``{EPOS Model and Ultra High Energy Cosmic Rays}'',
  \href{http://dx.doi.org/10.1016/j.nuclphysbps.2009.09.017}{{\em Nucl. Phys.
  Proc. Suppl.} {\bfseries 196} (2009) 102--105},
\href{http://arxiv.org/abs/0905.1198}{{\ttfamily arXiv:0905.1198 [hep-ph]}}.

\bibitem{Werner:2010aa}
K.~Werner, I.~Karpenko, T.~Pierog, M.~Bleicher, and K.~Mikhailov,
  ``{Event-by-Event Simulation of the Three-Dimensional Hydrodynamic Evolution
  from Flux Tube Initial Conditions in Ultrarelativistic Heavy Ion
  Collisions}'', \href{http://dx.doi.org/10.1103/PhysRevC.82.044904}{{\em Phys.
  Rev. C} {\bfseries 82} (2010) 044904},
  \href{http://arxiv.org/abs/1004.0805}{{\ttfamily arXiv:1004.0805 [nucl-th]}}.

\bibitem{Acharya:2019mzb}
{\bfseries ALICE} Collaboration, S.~Acharya {\em et~al.}, ``{Charged-particle
  production as a function of multiplicity and transverse spherocity in pp
  collisions at $\sqrt{s} =5.02$ and 13 TeV}'',
  \href{http://dx.doi.org/10.1140/epjc/s10052-019-7350-y}{{\em Eur. Phys. J.}
  {\bfseries C79} (2019) 857},
\href{http://arxiv.org/abs/1905.07208}{{\ttfamily arXiv:1905.07208 [nucl-ex]}}.

\bibitem{Tsallis:1987eu}
C.~Tsallis, ``{Possible Generalization of Boltzmann-Gibbs Statistics}'',
  \href{http://dx.doi.org/10.1007/BF01016429}{{\em J. Statist. Phys.}
  {\bfseries 52} (1988) 479--487}.
\url{https://doi.org/10.1007/BF01016429}.

\bibitem{Abelev:2013bla}
{\bfseries ALICE} Collaboration, B.~Abelev {\em et~al.}, ``{Multiplicity
  dependence of the average transverse momentum in pp, p-Pb, and Pb-Pb
  collisions at the LHC}'',
  \href{http://dx.doi.org/10.1016/j.physletb.2013.10.054}{{\em Phys. Lett.}
  {\bfseries B727} (2013) 371--380},
\href{http://arxiv.org/abs/1307.1094}{{\ttfamily arXiv:1307.1094 [nucl-ex]}}.

\bibitem{Loizides:2017ack}
C.~Loizides, J.~Kamin, and D.~d'Enterria, ``{Improved Monte Carlo Glauber
  predictions at present and future nuclear colliders}'',
  \href{http://dx.doi.org/10.1103/PhysRevC.97.054910,
  10.1103/PhysRevC.99.019901}{{\em Phys. Rev.} {\bfseries C97} (2018) 054910},
  \href{http://arxiv.org/abs/1710.07098}{{\ttfamily arXiv:1710.07098
  [nucl-ex]}}.
[erratum: Phys. Rev.C99,no.1,019901(2019)].

\bibitem{ABELEV:2013zaa}
{\bfseries ALICE} Collaboration, B.~Abelev {\em et~al.}, ``{Multi-strange
  baryon production at mid-rapidity in Pb$-$Pb collisions at
  $\sqrt{s_\mathbf{NN}}=2.76$ TeV}'',
  \href{http://dx.doi.org/10.1016/j.physletb.2014.05.052,
  10.1016/j.physletb.2013.11.048}{{\em Phys. Lett.} {\bfseries B728} (2014)
  216--227}, \href{http://arxiv.org/abs/1307.5543}{{\ttfamily arXiv:1307.5543
  [nucl-ex]}}.
[Erratum: Phys. Lett.B734,409(2014)].

\bibitem{Abelev:2012jp}
{\bfseries ALICE} Collaboration, B.~Abelev {\em et~al.}, ``{Multi-strange
  baryon production in $pp$ collisions at $\sqrt{s} = 7$ TeV with ALICE}'',
  \href{http://dx.doi.org/10.1016/j.physletb.2012.05.011}{{\em Phys. Lett.}
  {\bfseries B712} (2012) 309--318},
\href{http://arxiv.org/abs/1204.0282}{{\ttfamily arXiv:1204.0282 [nucl-ex]}}.

\end{thebibliography}\endgroup


\newpage
\appendix

%
%

\section{The ALICE Collaboration}
\label{app:collab}

\begin{flushleft}

\bigskip 

S.~Acharya$^{\rm 142}$, 
D.~Adamov\'{a}$^{\rm 97}$, 
A.~Adler$^{\rm 75}$, 
J.~Adolfsson$^{\rm 82}$, 
G.~Aglieri Rinella$^{\rm 34}$, 
M.~Agnello$^{\rm 30}$, 
N.~Agrawal$^{\rm 54}$, 
Z.~Ahammed$^{\rm 142}$, 
S.~Ahmad$^{\rm 16}$, 
S.U.~Ahn$^{\rm 77}$
% ...

\bigskip

\bigskip 

\textbf{\Large Affiliation Notes}

\bigskip 

% ...
$^{\rm VII}$ Also at: University of Kansas, Lawrence, Kansas, United States\\

\bigskip

\bigskip 

\textbf{\Large Collaboration Institutes}

\bigskip 

% ...
$^{16}$ Department of Physics, Aligarh Muslim University, Aligarh, India\\
% ...
$^{30}$ Dipartimento DISAT del Politecnico and Sezione INFN, Turin, Italy\\
% ...
$^{34}$ European Organization for Nuclear Research (CERN), Geneva, Switzerland\\
% ...
$^{54}$ INFN, Sezione di Bologna, Bologna, Italy\\
% ...
$^{75}$ Johann-Wolfgang-Goethe Universit\"{a}t Frankfurt Institut f\"{u}r Informatik, Fachbereich Informatik und Mathematik, Frankfurt, Germany\\
% ...
$^{77}$ Korea Institute of Science and Technology Information, Daejeon, Republic of Korea\\
% ...
$^{82}$ Lund University Department of Physics, Division of Particle Physics, Lund, Sweden\\
% ...
$^{97}$ Nuclear Physics Institute of the Czech Academy of Sciences, \v{R}e\v{z} u Prahy, Czech Republic\\
% ...
$^{142}$ Variable Energy Cyclotron Centre, Homi Bhabha National Institute, Kolkata, India\\
% ...

\bigskip 

\end{flushleft} 
  
\end{document}